\documentclass[a4paper,11pt]{article}
\usepackage{jcappub} % for details on the use of the package, please see the JINST-author-manual
\usepackage[utf8]{inputenc}
\usepackage[T1]{fontenc}
\usepackage[dvipsnames]{xcolor}
\usepackage{array}
\usepackage{graphicx}
\usepackage{adjustbox}
\usepackage{booktabs}
\usepackage{CJK}
\usepackage{enumitem}
\usepackage{soul}
\usepackage{amsfonts}

\def\Msunh{\mathinner{h^{-1}{\rm M}_{\odot}}}

\def\Mpch{\mathinner{h^{-1}{\rm Mpc}}}

\arxivnumber{2301.03278} % Only if you have one
\title{\boldmath MulGuisin, a Topological Network Finder and its Performance on Galaxy Clustering}

\author[a,b]{Young Ju,}
\author[a,b]{Inkyu Park,}
\author[a,b]{Cristiano G. Sabiu}
\author[c,d]{and Sungwook E. Hong}
\affiliation[a]{Department of Physics, University of Seoul, \\163 Seoulsiripdae-ro, Dongdaemun-gu, Seoul 02504, Republic of Korea}
\affiliation[b]{Natural Science Research Institute, University of Seoul, \\163 Seoulsiripdae-ro, Dongdaemun-gu, Seoul 02504, Republic of Korea}
\affiliation[c]{Korea Astronomy and Space Science Institute, \\776 Daedeok-daero, Yuseong-gu, Daejeon 34055, Republic of Korea}
\affiliation[d]{Astronomy Campus, University of Science and Technology, \\776 Daedeok-daero, Yuseong-gu, Daejeon 34055, Republic of Korea}

\emailAdd{icpark@uos.ac.kr}

\abstract{We introduce a new clustering algorithm, MulGuisin (MGS), that can identify distinct galaxy over-densities using topological information from the galaxy distribution.
This algorithm was first introduced in an LHC experiment as a Jet Finder software, which looks for particles that clump together in close proximity.
The algorithm preferentially considers particles with high energies and merges them only when they are closer than a certain distance to create a jet. 
MGS shares some similarities with the minimum spanning tree (MST) since it provides both clustering and network-based topology information. 
Also, similar to the density-based spatial clustering of applications with noise (DBSCAN), MGS uses the ranking or the local density of each particle to construct clustering.
In this paper, we compare the performances of clustering algorithms using controlled data and some realistic simulation data as well as the SDSS observation data, and we demonstrate that our new algorithm finds networks most efficiently and defines galaxy networks in a way that most closely resembles human vision.}

\begin{document}
\maketitle
\flushbottom

%===============================================================================
\section{Introduction}\label{sec:intro}
%===============================================================================

In the standard $\Lambda$CDM cosmology paradigm, structures in the universe grow in a hierarchical manner (e.g.,~\cite{White1978, Fall1980, Blumenthal1984}).
It means that smaller structures of matter start forming earlier, and the more massive structures form by the merging and accretion of smaller structures at later epochs of the universe.
Therefore, understanding cosmic structures in the universe at various scales is crucial for understanding the nature of our universe.
For example, numerous statistics of large-scale structures, such as topological analyses (e.g.,~\cite{Gott1986, Park1991, Park2010, Appleby2017, Appleby2018}), Alcock-Paczynski tests (e.g.,~\cite{Alcock1979, Ballinger1996, Li2014, hpark2019}), and small-scale redshift-space distortion (RSD; e.g.,~\cite{Sheth1996, DeRose2019, Tonegawa2020}), have been used to constraint cosmological parameters such as the matter density parameter ($\Omega_{\rm m}$) and the equation-of-state parameter of dark energy ($\omega_{\rm de}$).

While numerous statistics of cosmic structures have been used to understand the evolution and structure formation of our universe, the exact definition of cosmic structures remains unclear.
This is mainly because the matter distribution on a large scale is continuous, and therefore, there exists no specific discrete boundary for each structure.
Also, the membership for a certain structure might change if one considers other properties than just position, such as dynamics, mass, and so on (e.g.,~\cite{Serra2013, Gifford2013}).
Due to this ambiguity, numerous clustering algorithms have been proposed and used in the astronomical community.
For example, ref.~\cite{Knebe2011} compared the various properties of dark matter (DM) halos and subhalos found by 17 different halo-finding algorithms run on the same cosmological $N$-body simulation and found them to be in close agreement.

Galaxy clustering algorithms have played a crucial role in identifying galaxy clusters or super-clusters and exploring the large-scale structure of the universe, including filamentary structures. The most commonly utilized algorithms in astronomical research are the friends-of-friends (FoF; \cite{1985ApJ...292..371D}) and the minimum spanning tree (MST; \cite{boruvka1926jistem}) algorithms, which were introduced in the 1980s and have since become the standard for galaxy clustering. However, more recent algorithms, such as the halo-based algorithm \cite{Yang2007} and redMapper \cite{Rykoff2013}, have also been developed to address some of the limitations of the FoF and MST algorithms and are more directly applicable to observational data.

In recent years, with the rapid development of machine learning (ML) technology, clustering algorithms such as DBSCAN (Density-based Spatial Clustering of Applications with Noise; \cite{ester1996density}) have also been applied to galaxy clustering.
These clustering software, including the ML-based one, show comparable performance in galaxy clustering and produce consistent clustering results.
However, the results of clustering do not always represent the clusters that the human eye can find.
There are cases where a distribution clearly contains a cluster but it is not recognized as such by clustering algorithms, and there are cases where it is clearly divided into two clusters, visually, but appears as a single lump to the software.

As such, we have explored the possibility of developing an algorithm that creates galaxy clustering (or, networks) in a way that more closely resembles how the human eye and brain identify patterns. One approach we considered was to adapt jet-finding software used in high-energy particle physics research, with a particular focus on the MulGuisin (MGS) algorithm as a potentially suitable software for galaxy clustering.
MulGuisin (\begin{CJK}{UTF8}{mj}물귀신\end{CJK}) is a Korean word for a ghost that lives in water and is a figure that often appears in old Korean stories.
The MGS algorithm started with the idea that the ghosts hiding in the water could be found in the order of height by simply draining the water from the lake.

Initially, we copied the MGS software from the ATLAS (\emph{A Toroidal LHC Apparatus}) Jet-Finding library released in the early 1990s and developed it into a 3D galaxy network finding algorithm.
We then made several sample galaxy distributions to check the performance of MGS and compared the results to those produced by other standard clustering algorithms such as FoF, MST, and DBSCAN.
As a result, it was found that MGS has several important characteristics that other algorithms cannot demonstrate. The most important of these features is that as the linking length parameter for clustering increases, percolation occurs in all other algorithms, whereas in MGS, such a phenomenon does not occur even when a very large linking length is used. Therefore, MGS can be useful for finding large overdensity regions that are much broader than typical galaxy clusters by using very large linking lengths, which are not possible with other algorithms.

The networks identified by MGS are not necessarily all distinct, gravitationally bound collections of galaxies. Rather, MGS is a general method for identifying network structures in a spatial point set. In this work, we look at the performance of MGS at identifying clustered galaxies, compared with other algorithms, e.g., FoF, etc. In subsequent work, we will look in more detail at the physical properties of the MGS networks, e.g., halo mass function, and their usefulness in cosmological analysis.

The structure of this paper is as follows.
In section~\ref{sec:methods}, we introduce the MulGuisin clustering algorithm, as well as FoF, MST, and DBSCAN as benchmark clustering algorithms for comparison.
In section~\ref{sec:data}, we describe both controlled random data and realistic galaxy distribution data that we will use for the performance test.
We apply the above four clustering algorithms to the data and compare their performances in section~\ref{sec:results}, and we summarize our results in section~\ref{sec:conclusions}.

%===============================================================================
\section{Methods}\label{sec:methods}
%===============================================================================
In this section, we first introduce our MulGuisin clustering algorithm and introduce two clustering tools that are widely used in the field of astronomy and a newly developed clustering program through machine learning.
%-------------------------------------------------------------------------------
\subsection{MulGuisin network finding algorithm}
%-------------------------------------------------------------------------------
The MulGuisin (MGS) network finding algorithm was first introduced as a jet finder for the \emph{Large Hadron Collider} (LHC) physics in the ATLAS Collaboration \cite{Bosman:1998jfl}.
The algorithm is neither a variant of the conventional cone jet finding algorithm nor a variant of the $k_T$ clustering algorithm that is used in various collider experiments as the standard tools for finding jets. 
Although it has shown some improvements in jet reconstruction performance, such as optimized jet direction and jet energy resolution, but has not been used as a standard jet-finding tool for LHC experiments.

\begin{figure}[t]
\centering
\includegraphics[width=0.8\textwidth]{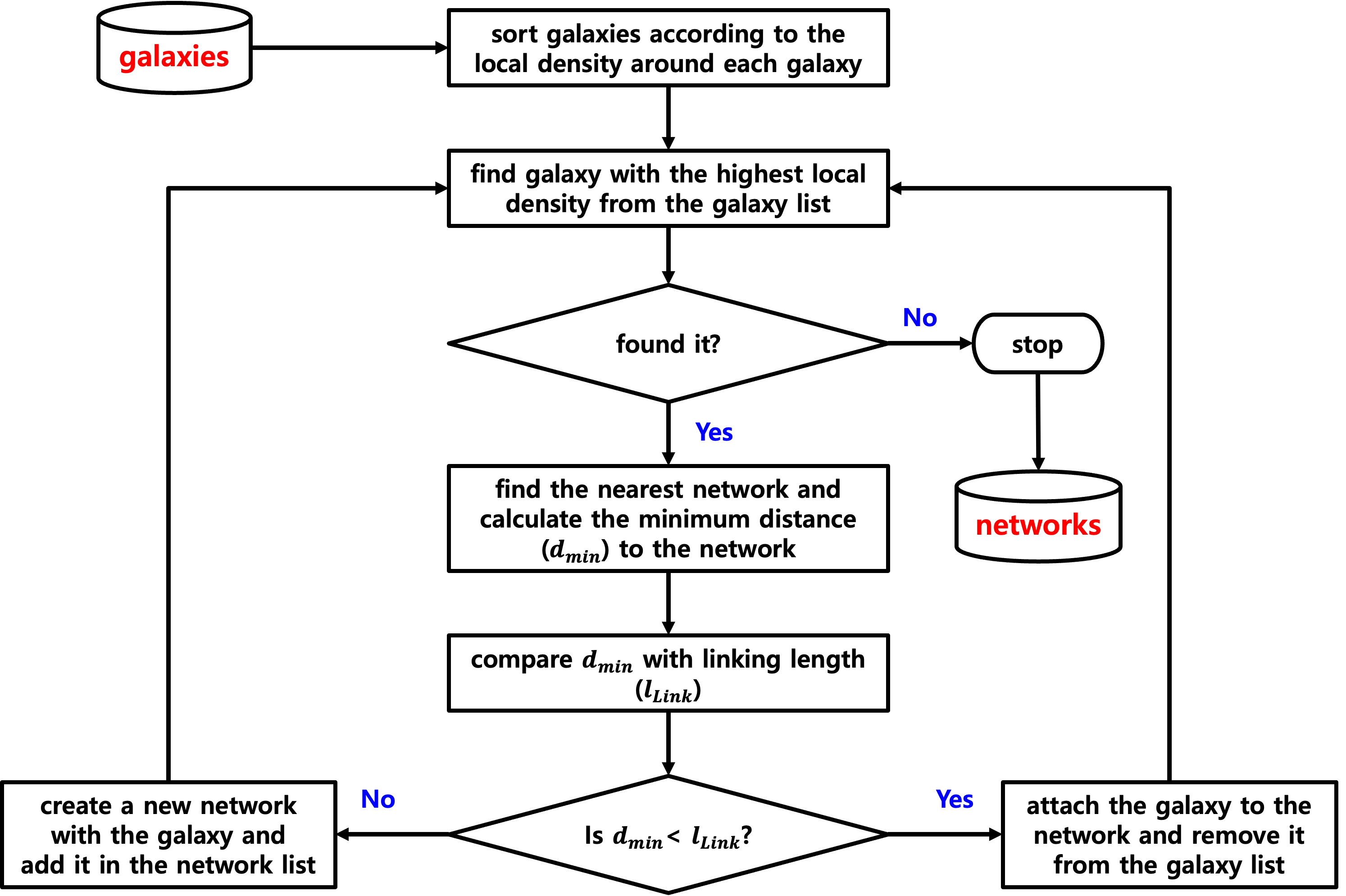}
\caption{Schematic flow chart to describe how MulGuisin (MGS) algorithm works}
\label{fig:flowchart}
\end{figure}

Figure~\ref{fig:flowchart} shows how the MGS algorithm works. 
The MGS algorithm first finds the point with the highest local number density from the input data and designates it as a network seed.
Then it finds the point with the second highest local number density and decides whether the point should belong to the first network or become a seed of a stand a new network.
This decision is made by checking how close the test point is to any neighboring networks, for which we introduce a parameter called the linking length ($\ell_{\rm Link}$).
That is, if the distance between the test point and the closest point in the network is less than this parameter, the test point is attached to the network, otherwise, it becomes the seed of a new network.
This parameter can be called resolution, which governs the separation of networks.
The algorithm then finds the next massive point and repeats the above process until there are no more points left to test.
At this stage, all points are assigned to networks or remain as isolated points.

\begin{figure}[t]
\centering
\includegraphics[width=0.8\textwidth]{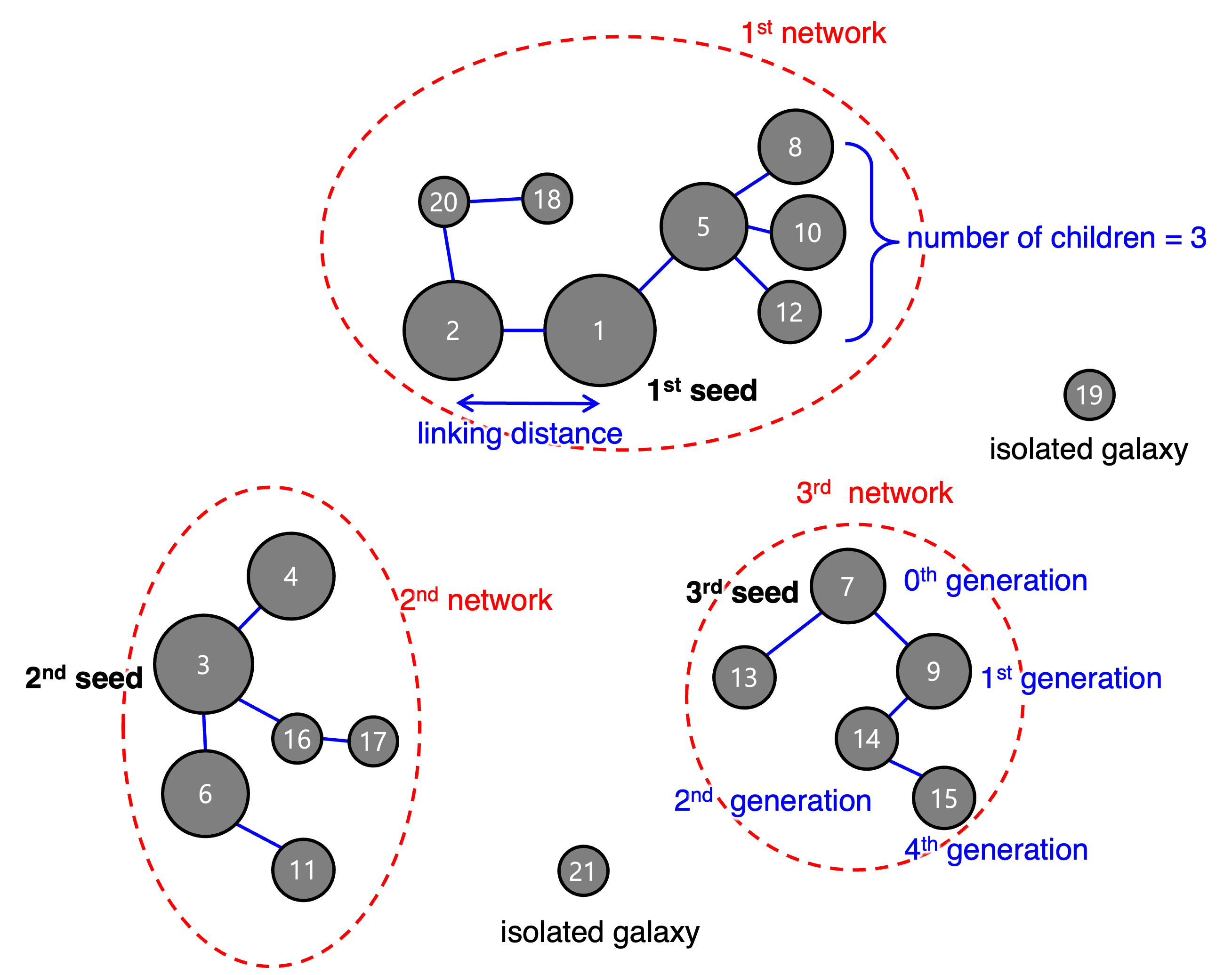}
\caption{Diagram showing how the MulGuisin (MGS) algorithm works to identify networks. Each gray circle represents a galaxy, and the size of the circle denotes its local number density, with the number specifying the galaxies ranking in descending order of density. 
In this specific example, 21 galaxies are grouped into 3 networks and 2 isolated galaxies.}
\label{fig:topo}
\end{figure}

Figure~\ref{fig:topo} is an illustration to explain how the MGS algorithm creates galaxy networks.
In the figure, the points are sorted in order according to their local number density which is estimated using a Voronoi tessellation.
According to this order, they become new network seeds or become attached to existing networks.
After going through the process, a network forms a tree-like structure that is sequentially connected according to the order of local number density.
The points in a network then form branches and nodes, and from the characteristic structure of such tree shape, one may able to study the topology of the galaxy network.

%- - - - - - - - - - - - - - - - - - - - - - - - - - - - - - - - - - - - - - - -
\subsection{Benchmark Algorithms}
%- - - - - - - - - - - - - - - - - - - - - - - - - - - - - - - - - - - - - - - -
In order to compare the performance of the MGS as a new clustering tool with the typical clustering algorithms, we select three benchmark algorithms, mostly based on their popularity in the astronomical community, mathematical clarity, and versatility.
They are the Friends-of-Friends (FoF), Minimum Spanning Tree (MST), and the Density-Based Spatial Clustering of Applications with Noise (DBSCAN).
Here we briefly introduce each package and describe how they make networks.
\footnote{Note that running our MGS algorithm from scratch may take a longer time than the above benchmark algorithms, especially when the number of data points is large.
We found that most of the MGS calculation time, for a large number of data points, is taken in constructing the Voronoi tesselation and calculating the local density for each point.
If we separate MGS into a density calculation and a tree building part, we find that the tree building takes a similar time to the benchmark algorithms.}
\subsubsection{Friends-of-Friends (FoF)}
The friends-of-friends (FoF) algorithm is a commonly used technique for identifying networks in astrophysical data \cite{Huchra1982, Tago2008, Duarte2014, Tempel2016}. 
This algorithm has a single free parameter, the linking length ($\ell_{\rm FoF}$), which determines the distance threshold for linking two data points. Points that are within this distance of each other are considered to be connected, and all connected points are grouped together into a single cluster.

One limitation of the FoF algorithm is that it can be difficult to choose an appropriate linking length. Different values of this parameter can result in networks of different shapes or numbers, making it challenging to determine the optimal value \cite{Tago2008}.\footnote{Note that the appropriate choice linking length for identifying DM halos from the DM particles in the $N$-body simulations is well known ($\ell_{\rm FoF} \simeq 0.2 \langle d_{\rm particle} \rangle$) \cite{More2011}. However, the optimal choice of linking length in general clustering problems is not well known.}
In this study, we use the \texttt{Halotools} implementation of the FoF algorithm \cite{Hearin_2017} to identify networks in our datasets by applying various $\ell_{\rm FoF}$.
Unless otherwise noted, we assume all FoF groups containing two or more members as networks.
 
\subsubsection{Minimum Spanning Tree (MST)} 

Galaxy data can be represented as a graph, with each galaxy represented as a node and the distance between two galaxies represented as an edge. The minimum spanning tree (MST) algorithm is a method for constructing a unique network from this data by connecting all nodes with minimum edges. Unlike other clustering algorithms, the MST does not require the use of a free parameter such as a linking length to construct the entire network.
However, the MST connects all nodes and may not produce networks 
with shapes that accurately reflect those of the original networks. 

Nevertheless, MST has been used in cosmology to study the large-scale structure of the universe \cite{Barrow1985, Krzewina1996, Naidoo2020}. In this study, we use the \texttt{MiSTree} package \cite{Naidoo2019} to construct MSTs from our galaxy data.   
Then, we find networks from the single MST tree by cutting nodes longer than the linking length ($\ell_{\rm MST}$).
Similar to the FoF case, we apply various values of $\ell_{\rm MST}$ and assume all tree segments containing two or more members as networks.

\subsubsection{Density-based Spatial Clustering of Applications with Noise (DBSCAN)}

The use of machine learning (ML) techniques is widespread in astronomy, as they enable the identification of patterns in data using algorithms. ML algorithms can be classified based on the type of data they are applied to, and one type, called unsupervised ML, is used with unlabeled data. Clustering algorithms, a subcategory of unsupervised ML algorithms, are used to group together data points with similar properties. One popular clustering algorithm is DBSCAN (density-based spatial clustering of applications with noise), which has been applied in a variety of contexts \cite{ester1996density, sander2017dbscan}.
    
DBSCAN is a density-based clustering algorithm that groups together data points based on their local density. 
In this algorithm, each network is identified by defining its core, which consists of high-density points within a certain distance.
The definition of core requires two free parameters, {\em min\_samples} and {\em eps}, which determine the minimum number of neighbors a point must have within a given radius in order to be considered as the core.
Then, other points that are directly reachable from some core points within {\em eps} are also considered part of the network, while other points are labeled as noise. 

In this study, we use the \texttt{scikit-learn} package \cite{scikit-learn} to implement the DBSCAN algorithm and identify networks in our data by applying various {\em eps} (or, the ``linking length'' in DBSCAN ($\ell_{\rm DBSCAN}$)).
Unless otherwise noted, we assume {\em min\_samples} $= 3$.

%- - - - - - - - - - - - - - - - - - - - - - - - - - - - - - - - - - - - - - - -
\subsection{A Simple 2D Toy Model Test}
\begin{figure}[t]
\centering
\includegraphics[width=0.8\textwidth]{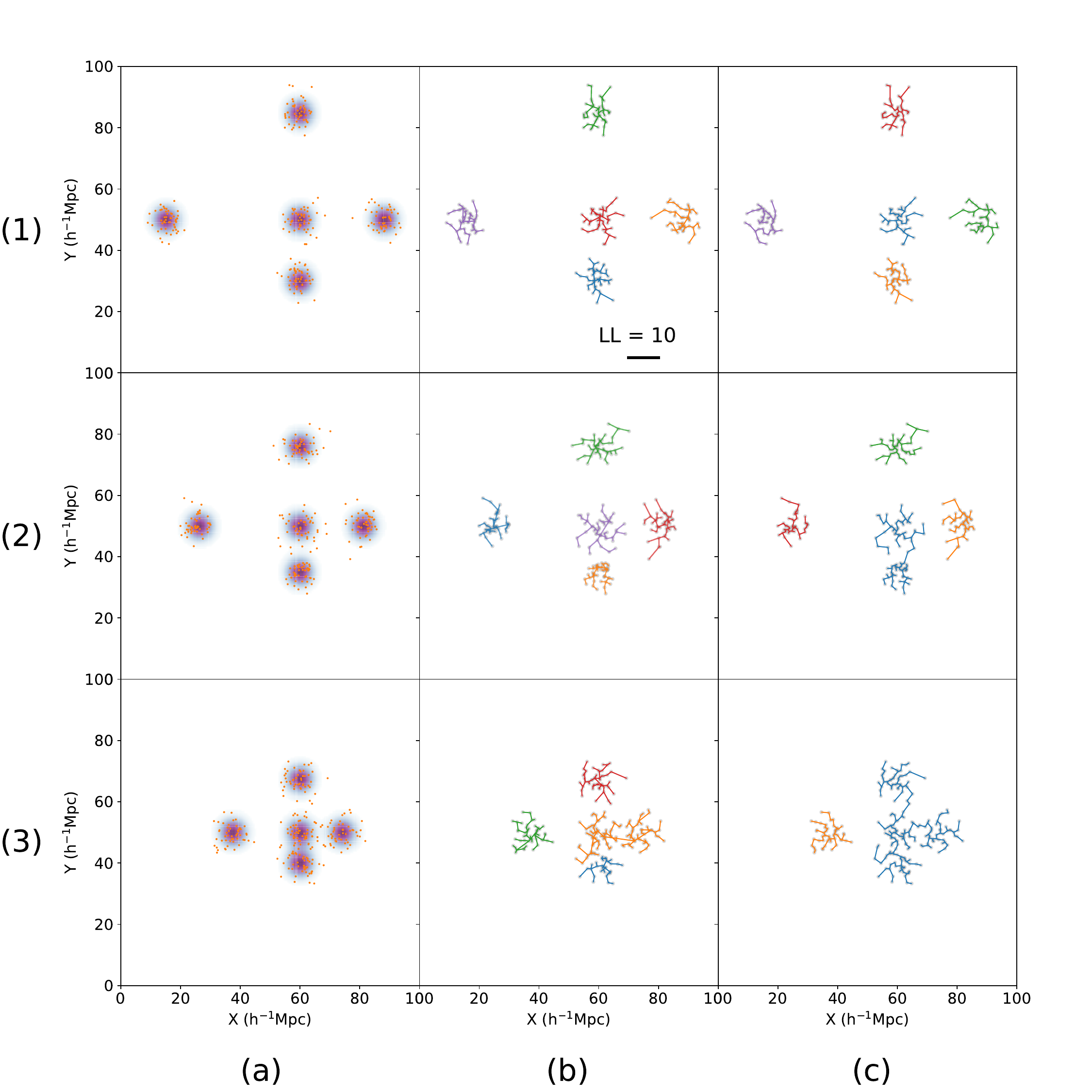}
\caption{A simple 2D toy model test of the MGS algorithm by comparing it with the MST algorithm. 
(a) Input distributions of 5 networks with different degrees of separation from each other ((1)--(3)). Background color denotes the galaxy number distribution we used for generating the galaxies. 
(b) Networks found by MGS and their tree structures. 
(c) Networks found by MST and their tree structures.
We used a linking length (LL) of $10\Mpch$ for both algorithms and we drew the size of the linking length in the top middle figure.}
\label{fig:2d-sample}
\end{figure}

%- - - - - - - - - - - - - - - - - - - - - - - - - - - - - - - - - - - - - - - -
To see how the shape of the networks generated by the MGS algorithm differs from the results of other clustering algorithms, we created simple simulation data and compared the results.
We first assume that there are 5 networks in 2D space, and consider the case where each network contains 50 galaxies equally.
The galaxies are distributed within the network according to a Gaussian function with a width of $10\Mpch$.
The coordinates of the two-dimensional space span from 0 to $100\Mpch$ on both the X- and Y-axes, and the position of each network is set to have three different distributions, from far away from each other to all close together, as shown in figure~\ref{fig:2d-sample} column (a) in rows (1), (2) and (3).

As shown in figure~\ref{fig:2d-sample}, both the MGS and MST algorithms correctly find 5 networks when the distances among the networks are sufficiently far apart.
However, when the networks get closer together, MST cannot differentiate between the networks and starts recognizing them as one big network.
Even for the cases where networks are attached to each other as shown in figure~\ref{fig:2d-sample} (3), MGS still recognizes four among five true networks like the human eyes can distinguish each network, whereas MST recognizes 4 adjacent networks as one huge network.
These differences can create serious differences in results when studying the number and mass distributions of networks.

In the next section, we will compare the performance of MGS and other algorithms with more realistic 3D data.
%===============================================================================
\section{Data}\label{sec:data}
%===============================================================================

Our final goal is to use the MGS algorithm for galaxy clustering or other large-scale structure finding for the cosmology study.
However, since some inconsistencies exist between various clustering algorithms for finding networks or other large-scale structures 
 (e.g., see \cite{Knebe2011} and references therein), we cannot compare the MGS networks found in the realistic data with their ``truth''.

Therefore, we apply two types of data sets in this section to compare the performance between MGS and other benchmark algorithms.
The first sets, called the ``controlled random data'' (D1--D3), are those that we design all properties of networks, including their positions and member galaxy distributions.
Since we already know the true information of each network, we can test which algorithms predict the true networks better in which conditions.
The next sets, called the ``realistic data'' (D4), are the observational and simulation data sets of galaxies around $z \simeq 0$, and we focus on comparing the properties of predicted networks in each algorithm.
Table~\ref{tab:data} summarizes the data sets we use in this work.

\begin{table}[t]
\centering
\begin{tabular}{l p{0.85\textwidth}}
\toprule
Data set & Description \\
\midrule
 D1-LD          &  50 randomly positioned networks, each of which contains 100 galaxies randomly spread by the 3D Gaussian distribution with standard deviation $\sigma = 1\Mpch$. The total number of galaxies is 5,000. \vspace{6pt} \\
 D1-HD         &  Same as D1-LD, but with the greater standard deviation $\sigma = 10\Mpch$. \\
 \midrule
 D2-NA   &  Same as D1-HD, but the number of galaxies in each network follows an exponential random distribution. 
 The total number of galaxies is 7,041.  \vspace{6pt} \\
 D2-LA  &  Same as D2-NA, but adding uniformly randomly distributed noisy galaxies to the entire box to increase the total galaxy number density 1.5 times of D2-NA.The total number of galaxies is 12,041. \vspace{6pt} \\
 D2-HA &  Same as D2-LA, but adding more noisy galaxies so that the total galaxy number is twice D2-NA.
 The total number of galaxies is 17,041. \\
 \midrule
 D3-HOD & 500 randomly positioned networks with a mass distribution similar to the Press-Schechter mass function. The number of galaxies for each network follows HOD \cite{Kravtsov2004} for massive halos ($M \geq 10^{13} \Msunh$). The galaxies are spread by the NFW profile with the concentration parameter of 10.
 The total number of galaxies is 65,460. \\
 \midrule
 D4-SDSS & Volume-limited sample of the KIAS-VAGC \cite{choi2010} with absolute $r$-band magnitude $\mathcal{M}_r - 5 \log h < -20$. \vspace{6pt} \\
 D4-HR4 & Four lightcone data of mock galaxy catalogs from the Horizon Run 4 simulation \cite{kim2015,hong2016} with a similar condition to D4-SDSS. \\
 \bottomrule
\end{tabular}
\caption{Name and description of galaxy data sets that we use in this analysis. The box size of all controlled data (D1--D3) is $(200\Mpch)^{3}$.}
\label{tab:data}
\end{table}

\subsection{Controlled Random Data (D1--D3)}\label{sec:data_random}

\subsubsection{Different Spatial Dispersion (D1)}

\begin{figure}[t]
\centering
\includegraphics[width=\textwidth]{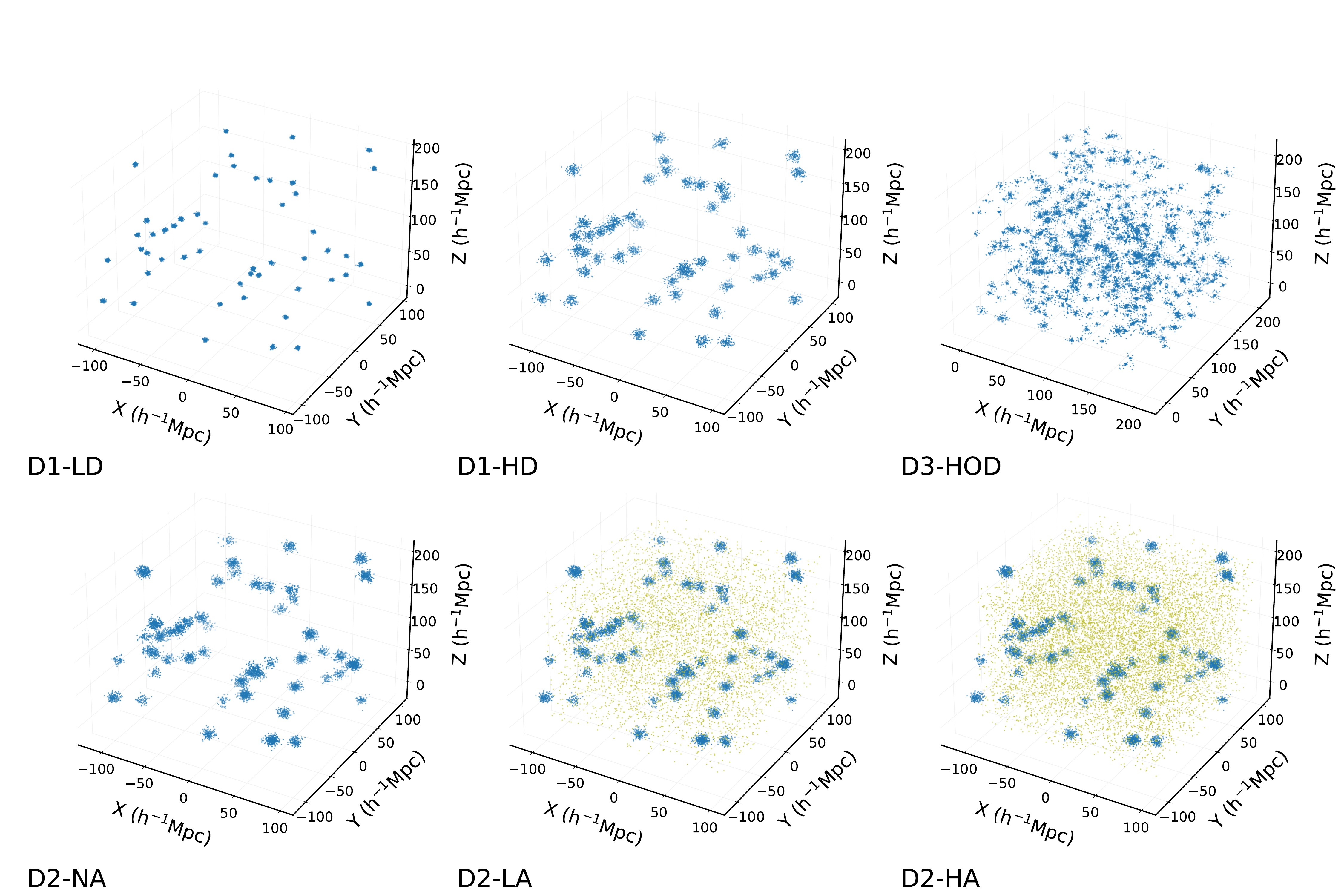}
\caption{Three-dimensional galaxy distributions of controlled data sets used in this paper.
Top: D1-LD(left), D1-HD(middle), and D3-HOD(right).
Bottom: D2-NA(left), D2-LA(middle), and D2-HA(right), with noisy additional galaxies shown as yellow dots.
See table~\ref{tab:data} for details.}
\label{fig:controlled_dist}
\end{figure}

We use controlled, simulated data to evaluate the performance of the MGS algorithm in comparison to other clustering algorithms. These data are generated randomly and allow us to control the shape and distribution of networks to test the algorithms under different conditions. The first set of data consists of 100 galaxies per network and 50 networks within a 3-dimensional cubic volume of space with a side length $200\Mpch$. The network center positions are chosen randomly, and the galaxies in each network are distributed according to a Gaussian distribution with a variable standard deviation ($\sigma$) that controls the spatial dispersion. The D1-LD data has a low spatial dispersion ($\sigma = 1\Mpch$), leading to well-separated networks, while the D1-HD data has a higher spatial dispersion ($\sigma = 10\Mpch$), resulting in networks that are closer together. 
Upper left and middle panels of figure~\ref{fig:controlled_dist} show the distribution of galaxies in the D1-LD and D1-HD data sets.

\subsubsection{Additional Noisy Galaxies (D2)}

We generate additional controlled data sets that are similar to D1 but with slightly different characteristics. We again place 50 network centers at the same positions as D1, but this time we use an exponential distribution to generate a variable number of galaxies for each network:
\begin{equation}
P(N_{\rm gal}) = \left\{ \begin{array}{ll}
\displaystyle{\frac{1}{\Delta N} \exp \left[ -\frac{N_{\rm gal} - N_0}{\Delta N} \right]} & \textrm{ if } N_{\rm gal} > N_0 \\
0 & \textrm{ otherwise} \\
\end{array} \right. ~.
\end{equation}
Here, we set $N_0$ and $\Delta N$ as 50 and 100, respectively, so that the minimum number of galaxies per network and the total number of galaxies roughly match with D1.
The galaxies are spatially distributed according to a Gaussian function centered on the network's center with a standard deviation of $\sigma=10\Mpch$, as was done for D1-HD, and we call this new controlled data D2-NA. The lower left panel of figure~\ref{fig:controlled_dist} shows the distribution of galaxies in the D2-NA data. The total number of galaxies in this data set is 7,041. 

In addition to D2-NA, we introduce two more data sets created by adding unclustered galaxies, which are sampled uniformly in the entire box.  
We add these  `noisy' galaxies so as to test how the algorithms are affected by the background density.  
The lower middle panel of figure~\ref{fig:controlled_dist} shows the D2-LA data, where we add 5,000 galaxies (yellow dots) to increase the galaxy number density by 1.5 times compared to D2-NA. 
On the other hand, the lower right panel shows the D2-HA data, where we add 10,000 galaxies to make the total galaxy number density twice that of D2-NA.

%- - - - - - - - - - - - - - - - - - - - - - - - - - - - - - - - - - - - - - - -
\subsubsection{HOD-based Mock Galaxies (D3-HOD)}
%- - - - - - - - - - - - - - - - - - - - - - - - - - - - - - - - - - - - - - - -
We generate a third set of controlled, simulated data to create a more complex environment for testing the performance of the MGS algorithm. We use an analytic formula to model the distribution of galaxies in this data set. 
First, we create 500 network center positions by sampling uniform random distribution within a $(200\Mpch)^3$ box. 
Then, we obtain the normalized version of Press-Schechter halo mass function at $z = 0$ \cite{Press1974}, with a concordance $\Lambda$CDM cosmology to the {\em Planck} 2015 data \cite{Planck2016}, for massive halos $M_{\rm halo} > 10^{13}\Msunh$ using the \texttt{Colossus} package \cite{Diemer_2018}.
We then obtain masses for each of the 500 networks 
by randomly sampling for the mass function.\footnote{Note that neither the positions nor the mass distribution of networks in D3-HOD follow the estimation from the standard cosmology. However, here we focus only on providing complex environments, and therefore, such differences do not affect our motivation. See section~\ref{sec:data_real} for realistic data sets instead.}

We then use this information to generate a distribution of the number of galaxies using a halo occupation distribution (HOD) model. The mean halo occupation is typically assumed to follow a power law at massive halo masses \cite{Berlind2002, Kravtsov2004}: 
\begin{equation}
N_{\mathrm{avg}}(M_{\rm network}) = \left\{ \begin{array}{ll} \displaystyle{\left( \frac{M_{\rm network}}{M_1} \right)^\alpha} & \textrm{if } M_{\rm halo} > M_{\mathrm{min}} \\ 
0 & \textrm{otherwise}  \\ \end{array} \right. ~,
\end{equation}
where $\alpha$, $M_{\rm min}$, and $M_1$ correspond to the power-law index, cutoff halo mass where halo cannot contain galaxies, and the mass scale containing a single galaxy at the given condition of galaxy sample.
Here, we use $\alpha = 0.87$ and $M_{\mathrm{min}} = 10^{13} \Msunh$ by following ref.~\cite{Kravtsov2004}.
We set $M_1 = 10^{11} \Msunh$ so that the minimum number of galaxies for each network is set as 50.
Also, for simplicity, we calculate the actual number of galaxies at each network by applying the ceiling to $N_{\rm avg}$.

Next, we use the \texttt{Colossus} package to create an Navarro-Frenk-White (NFW) profile \cite{Navarro1996}
\begin{equation}
\rho(x) = \frac{M_{\rm network}}{4 \pi R_{\rm vir}^3} \left[ \left\{ \ln (1+c_{\rm s}) - \frac{c_{\rm s}}{1+c_{\rm s}} \right\} x (x + c_{\rm s}^{-1})^2 \right]^{-1} ~,
\end{equation}
where $x \equiv r/R_{\rm vir}$.
The concentration parameter for the NFW profile $c_{\rm s}$ is fixed as 10, and the virial radii $R_{\rm vir}$ is determined by the network mass accordingly.
We then randomly distribute the galaxies according to this profile, and the resulting data set consists of 65,460 galaxies.
The upper right panel of figure~\ref{fig:controlled_dist} shows the distribution of galaxies in D3-HOD.

\subsection{Realistic Data: SDSS \& Horizon Run 4 (D4)}\label{sec:data_real}

In the previous subsection, we described a set of controlled data catalogs for which we can carefully control the properties of the networks.
Such data are useful for testing the performance of MGS over other benchmark algorithms by comparing the properties of identified networks with the input truth. 
However, the true distribution of galaxies in the universe differs from these controlled random data in the following ways.
First, unlike those in the controlled random data with low noise levels, the boundaries of clusters in the universe are often not clearly defined (e.g.,~\cite{Serra2013,Gifford2013}).
Also, the spatial distribution of galaxies in each cluster may not follow spherical symmetry (e.g., \cite{2013SSRv..177..155L} for a good review).
Furthermore, the redshift-space distortion elongates spherical clusters in real space, which may require that we separate linking lengths between the radial and tangential directions \cite{Farrens:2011eb, Tempel2016}.

Therefore, it is necessary to adopt a realistic galaxy distribution for a fair performance test of the MGS algorithm.
However, unlike in the case of the controlled random data where we know the answer, we may only study the difference between the network properties from the MGS and other benchmark algorithms.
Here we use observational data and four corresponding sets of mock simulation data --- the volume-limited KIAS-Value Added Galaxy Catalog (KIAS-VAGC) of the \emph{Sloan Digital Sky Survey} (SDSS) Main Galaxy Sample with $r$-band absolute magnitude $\mathcal{M}_r - 5 \log h < -20$ \cite{choi2010} and the lightcone mock galaxy samples from the Horizon Run 4 simulation \cite{kim2015, hong2016}.

%- - - - - - - - - - - - - - - - - - - - - - - - - - - - - - - - - - - - - - - -
\subsubsection{Volume-limited KIAS-VAGC (D4-SDSS)} \label{sec:data_real_sdss}
%- - - - - - - - - - - - - - - - - - - - - - - - - - - - - - - - - - - - - - - -
The KIAS Value-Added Galaxy Catalog (KIAS-VAGC; \cite{choi2010}) is an upgraded version of the New York University Value-Added Galaxy Catalog (NYU-VAGC; \cite{2005AJ....129.2562B}), which is part of the \emph{Sloan Digital Sky Survey} (SDSS) Data Release 7 \cite{2009ApJS..182..543A}, by adding some missing redshifts  
to improve spectroscopic completeness.\footnote{The completeness of spectroscopic redshift for our sample (D4-SDSS) is 94.2\%. The remaining 5.8\% had suffered the fiber collision(see the main text), and the spectroscopic redshifts of their nearest neighbor galaxies are assigned to them instead.}
This catalog has been widely used in numerous studies, including cosmic voids statistics \cite{Pan2012, Hoyle2012}, largest structures of the universe \cite{Park2012}, the fraction of barred galaxies \cite{Lee2012a}, and the properties of active galactic nuclei (AGN; \cite{Lee2012b, Hwang2012, Bae2014}).

\begin{figure}[t]
\centering
\includegraphics[height=0.228\textwidth]{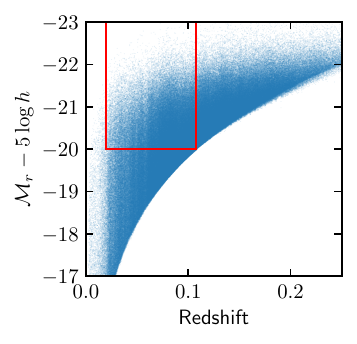}
\includegraphics[height=0.228\textwidth]{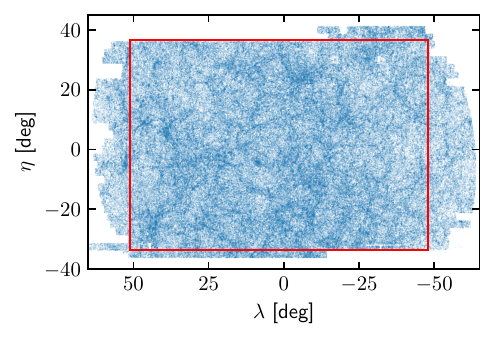}
\includegraphics[height=0.228\textwidth]{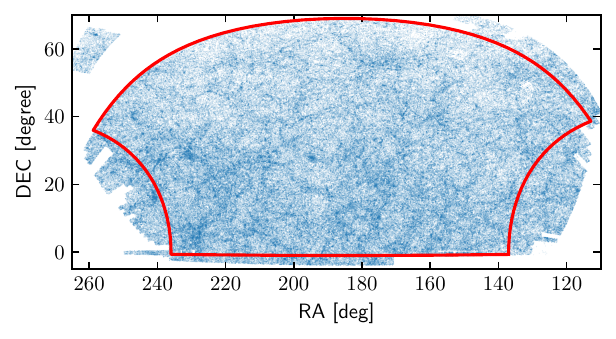}
\caption{Selection of the volume-limited sample of the KIAS Value-Added Galaxy Catalog (KIAS-VAGS) used in this study (red boxes).
Left: Volume-limited selection in the redshift vs. absolute $r$-band magnitude plane with $\mathcal{M}_r - 5 \log h < -20$.
Middle \& Right: Sky selection shown in both SDSS Survey coordinates $(\eta, \lambda)$ (middle) and (RA, DEC) coordinates (right).}
\label{fig:sdss}
\end{figure}

Most of the KIAS-VAGC galaxies were observed with the apparent $r$-band magnitude limit $r = 17.6$.
It means that, in terms of absolute magnitude, the catalog contains less bright galaxies at lower redshifts, while only very bright galaxies could be seen at higher redshifts.
Therefore, for a fair comparison between galaxies over a wide redshift range, we apply a ``volume-limited'' selection by selecting galaxies brighter than a certain absolute $r$-band magnitude \cite{Choi2010b}.
Here, we use $\mathcal{M}_r - 5 \log h < -20$.
By combining with the given apparent $r$-band magnitude limit, such absolute magnitude cutoff naturally provides the upper redshift bound of our volume-limited sample ($z < 0.107$; left panel of figure~\ref{fig:sdss}).
We also apply the lower redshift bound $z > 0.02$, by considering the incompleteness of the galaxy sample below the given redshift.

In addition to the volume-limited selection in the redshift-magnitude plane, we also apply a sky selection for simplification.
Specifically, instead of using spherical coordinates with right ascension (RA) and declination (DEC) $(\alpha, \delta)$, we select galaxies with the SDSS Survey coordinate ($\eta$, $\lambda$), defined by
\begin{align}
    \eta &\equiv \tan^{-1} \left[ \frac{\sin \delta}{\cos \delta \cos (\alpha - \alpha_0 )} \right] - \delta_0 \\
    \lambda &\equiv \sin^{-1} \left[ \cos \delta \sin (\alpha - \alpha_0 )\right] \, .
\end{align}
Here, $(\alpha_0, \delta_0) = (185^\circ, 32.5^\circ)$ is the RA/DEC sky position where $(\eta, \lambda) = (0, 0)$.
We select the rectangular sky region $-33.5^\circ < \eta < 36.5^\circ$ and $-48^\circ < \lambda < 51^\circ$, in order to maximize the sky area with simple geometry and to avoid issues arising from a complicated boundary (middle and right panels of figure~\ref{fig:sdss}).

%- - - - - - - - - - - - - - - - - - - - - - - - - - - - - - - - - - - - - - - -
\subsubsection{Horizon Run 4 (D4-HR4)}\label{sec:data_real_hr4}
%- - - - - - - - - - - - - - - - - - - - - - - - - - - - - - - - - - - - - - - -
The Horizon Run 4 simulation (HR4; \cite{kim2015}) is an extremely large cosmological $N$-body simulation that uses $6,300^3$ DM particles within a periodic cube with a comoving volume $V = (3.15~h^{-1} {\rm cGpc})^3$.
It assumes a vanilla $\Lambda$CDM cosmological model in concordance with the \emph{Wilkinson Microwave Anisotropy Probe} (WMAP) 5th-year result \cite{dunkley2009}.
Among 2,001 timesteps between $z = 100$ to 0, 75 coarse timesteps with mean time difference $\Delta t = 0.18~{\rm Gyr}$ are chosen between $z = 12$ to 0 to build a merging tree of FoF halos. 
The FoF linking length is $0.2$ times the particle mean separation, and we identify halos only whose mass is greater than $M_{\rm halo}^{\min} = 2.7 \times 10^{11} \Msunh$.

The mock galaxies are then produced by so-called the most bound halo particle (MBP)-galaxy abundance matching method \cite{hong2016}.
We find MBPs for all halos in the merging tree and adopt their positions and peculiar velocities as those of corresponding mock galaxies.
The ``mass'' of mock galaxies, which is used as a proxy of stellar mass or luminosity, is defined as the mass of their hosting halos.
For satellite halos, we identify their MBPs at the timestep just before the infall event and trace them until they are totally absorbed toward their central halo by tidal disruption. 
For estimating the tidal disruption timescale $t_{\rm merge}$), we adopt a modified model of ref.~\cite{jiang2008},
\begin{equation}
\frac{t_{\rm merge}}{t_{\rm dyn}} = \frac{(0.94 \epsilon^{0.60} + 0.60)/0.86}{\ln [1 + (M_{\rm host}/M_{\rm sat})]} \left( \frac{M_{\rm host}}{M_{\rm sat}} \right)^\alpha ~,
\end{equation}
where $\epsilon$, $M_{\rm host}$, $M_{\rm sat}$, $t_{\rm dyn}$ are the circularity of the satellite's orbit, the mass of central and satellite halos, and the orbital period of virialized objects, respectively.
We adopt $\alpha = 1.5$ for a better match of the galaxy two-point correlation function (2pCF) at scales less than $1\Mpch$ at a given spatial resolution of the HR4 \cite{Zehavi2011, hpark2019}.
Then the mass of survived satellite galaxies is defined as the mass of their hosting halos just before the infall.

After producing snapshot mock galaxy catalogs for coarse timesteps, we then produce lightcone mock galaxy catalogs up to $z = 1.5$.
The all-sky lightcone DM particle data of the HR4 were created during the simulation by stacking the comoving shells at the corresponding redshifts.
Then, we compare the IDs of the galaxy MBPs at each coarse timestep snapshot and those of DM particles at the lightcone data with the coarse comoving shells.
If the MBP ID of a given mock galaxy matches that of a particle in the lightcone data, we assign a galaxy in the lightcone data.
Here, we adopt the position and peculiar velocity from the particle at the lightcone data, while the galaxy ``mass'' comes from the mock galaxy at the nearest snapshot.

After creating the all-sky lightcone mock galaxy catalog, we cut it in a similar way to the volume-limited KIAS-VAGC sample.
First, we apply the redshift space distortion (RSD) for each mock galaxy for a fair comparison with observation, by using real-space positions and peculiar velocities.
Then, we apply the same redshift range $0.02 < z < 0.107$ and set the lower bound of galaxy ``mass,'' so that the galaxy number density of the HR4 lightcone data is identical to that of KIAS-VAGC.
After that, we create four non-overlapping subsets from it with the same angular geometry as our SDSS Survey coordinate selection.

During the analysis, we found that the fiber collision in the fiber-fed spectroscopic observations affects various clustering statistics \cite{Zehavi2002, Guo2012, Reid2014, Tonegawa2020}. 
Therefore, for a fair comparison, our HR4 mock galaxy catalogs also need to follow the same fiber collision condition as the KIAS-VAGC.
To do so, we select pairs of mock galaxies whose angular distance is less than 55~arcseconds and keep only one from each pair by random selection. Because SDSS observations were partially overlapping, some close pairs have both redshifts. In order to reflect this, we only fiber-collide 60\% of the close pairs.

%===============================================================================
\section{Results}\label{sec:results}
%===============================================================================

We evaluate the performance of the MGS algorithm and other relevant methods using three controlled datasets and observational data. Four distinct algorithms are executed with varying linking lengths to determine the number of networks.
However, assessing these algorithms solely based on the number of networks identified is insufficient. Even if an algorithm successfully recovers the correct network count, it remains uncertain whether the identified networks genuinely correspond to the original networks.

To account for this property and ensure that the estimated networks from the algorithms align with the original networks, we employ the Adjusted Rand Index (ARI; \cite{morey1984, hubert1985}), a modification of the Rand Index (RI; \cite{rand1971}). First, RI measures the similarity between two clustering outcomes by counting pairs of elements under specific conditions and is defined as
\begin{equation}
\mathrm{RI} = \frac{\mathrm{TP} + \mathrm{TN}}{\mathrm{TP} + \mathrm{FP} + \mathrm{FN} + \mathrm{FP}}.
\end{equation}
Here, TP (true positives) represents the number of pairs where the algorithm accurately identifies original network members. TN (true negatives) denotes the number of pairs where the algorithm correctly distinguishes non-members. FP (false positives) stands for the number of pairs where the algorithm incorrectly classifies non-members as members. FN (false negatives) signifies the number of pairs where the algorithm wrongly classifies members as non-members.

The ARI is then defined as
\begin{equation}
    \mathrm{ARI} = \frac{\mathrm{RI} - \mathbb{E}[\mathrm{RI}]}{1 - \mathbb{E}[\mathrm{RI}]}.
\end{equation}
Here, $\mathbb{E}[\mathrm{RI}]$ is an expectation value of the RI upon permutation of clustering indices.
An ARI value close or less than zero indicates that the algorithm is unable to recover the original network, while a value close or equal to one signifies efficient recovery of the original network.
We assess the performance of the algorithms using ARI as a function of linking length. Here, we set the minimum value of ARI as zero.

We use the three controlled data sets and observation data to evaluate the performance of the MGS algorithm and compare it to other clustering algorithms. We run each of the four algorithms with different values of the linking length and calculate the ARI of results identified by each algorithm. We repeat this process for a range of linking lengths and analyze the trends in the number of networks identified by each algorithm. This allows us to assess the sensitivity of the algorithms to the choice of linking length and to compare their performance in identifying networks in the different data sets.
%-------------------------------------------------------------------------------
\subsection{Results with Controlled Data}\label{sec:results_control}
%-------------------------------------------------------------------------------

\begin{figure}[t]
\centering
\includegraphics[width=0.6\textwidth]{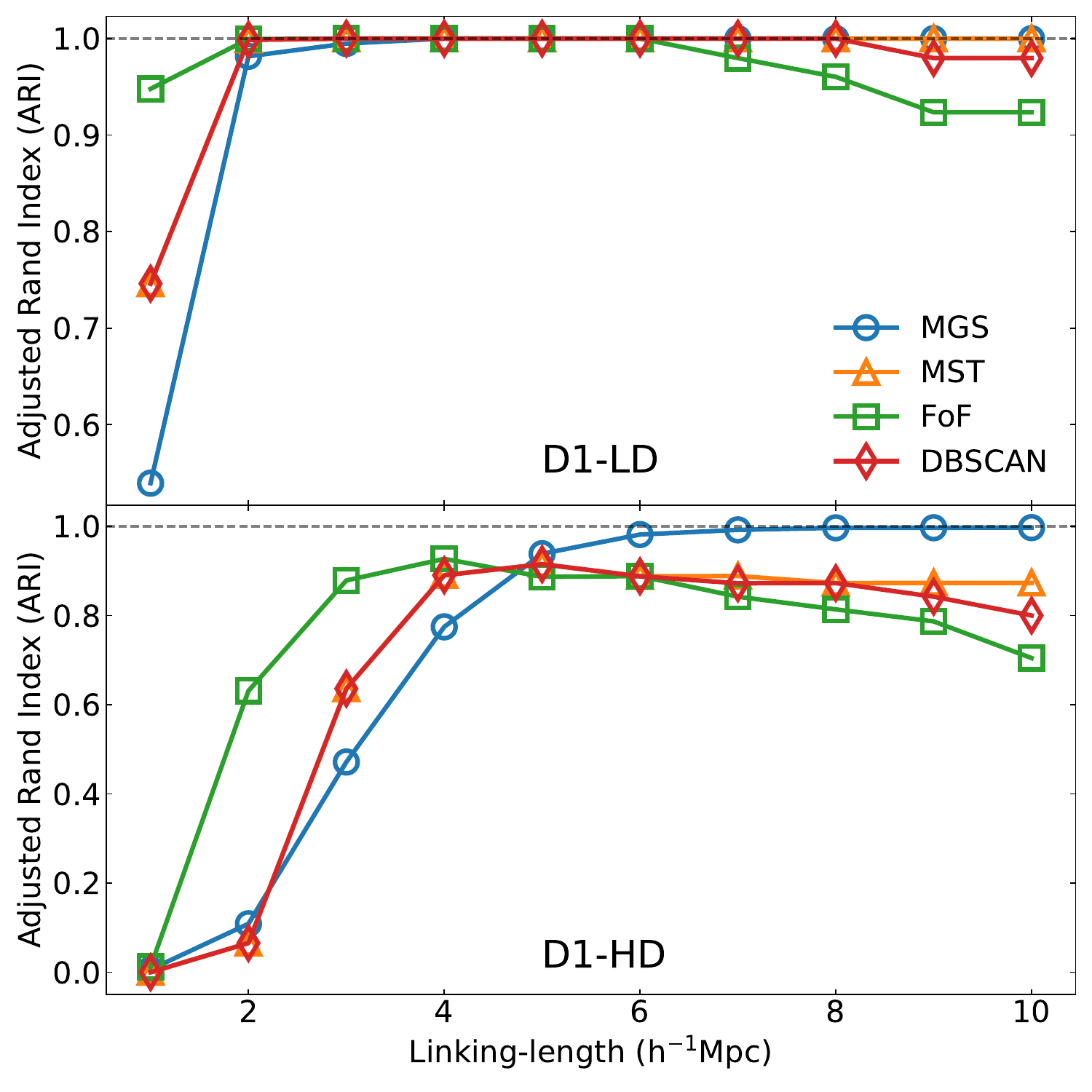}
\caption{The Adjusted Rand Index (ARI) as a function of linking length for D1-LD (top panel) and D1-HD (bottom).
Each of the 4 clustering algorithms is indicated using different colors and symbols. 
Note that MST and DBSCAN show considerable overlap.
}
\label{fig:4alg1}
\end{figure}

Figure~\ref{fig:4alg1} shows the ARI obtained by each algorithm as a function of the linking length for the controlled data set 1. The top panel shows the results for the D1-LD data, which consists of well-separated networks.
All four algorithms perform well in identifying the networks, but the MGS algorithm stands out for its ability to accurately identify the correct number of networks.
In particular, for large linking lengths, the FoF and DBSCAN algorithms identify fewer than 50 networks, resulting in a lower ARI, because they connect neighboring networks and merge them into a single network.

The bottom panel of figure~\ref{fig:4alg1} shows the results for the D1-HD data, which has a higher level of spatial dispersion and some networks that are close to each other. For smaller linking lengths, the algorithms identify many more networks than the true input value of 50. If the linkage length is too small the algorithm cannot connect all the galaxies in each network, and only a few galaxies become a new network.
That is, one network is artificially divided into many individual segments, and consequently, the algorithms have very low ARI values. As the linking length increases, the behavior of the algorithms becomes more distinct. The MGS algorithm continues to accurately identify the correct number of networks, while the other algorithms identify fewer networks due to the merging of originally separate networks. 
This is the opposite reason for why the ARI is low at larger linking lengths as compared to lower linkage lengths. The MGS algorithm is able to track the structure of the networks and identify their boundaries, leading to more accurate results in this type of data.

\begin{figure}[t]
\centering
\includegraphics[width=0.6\textwidth]{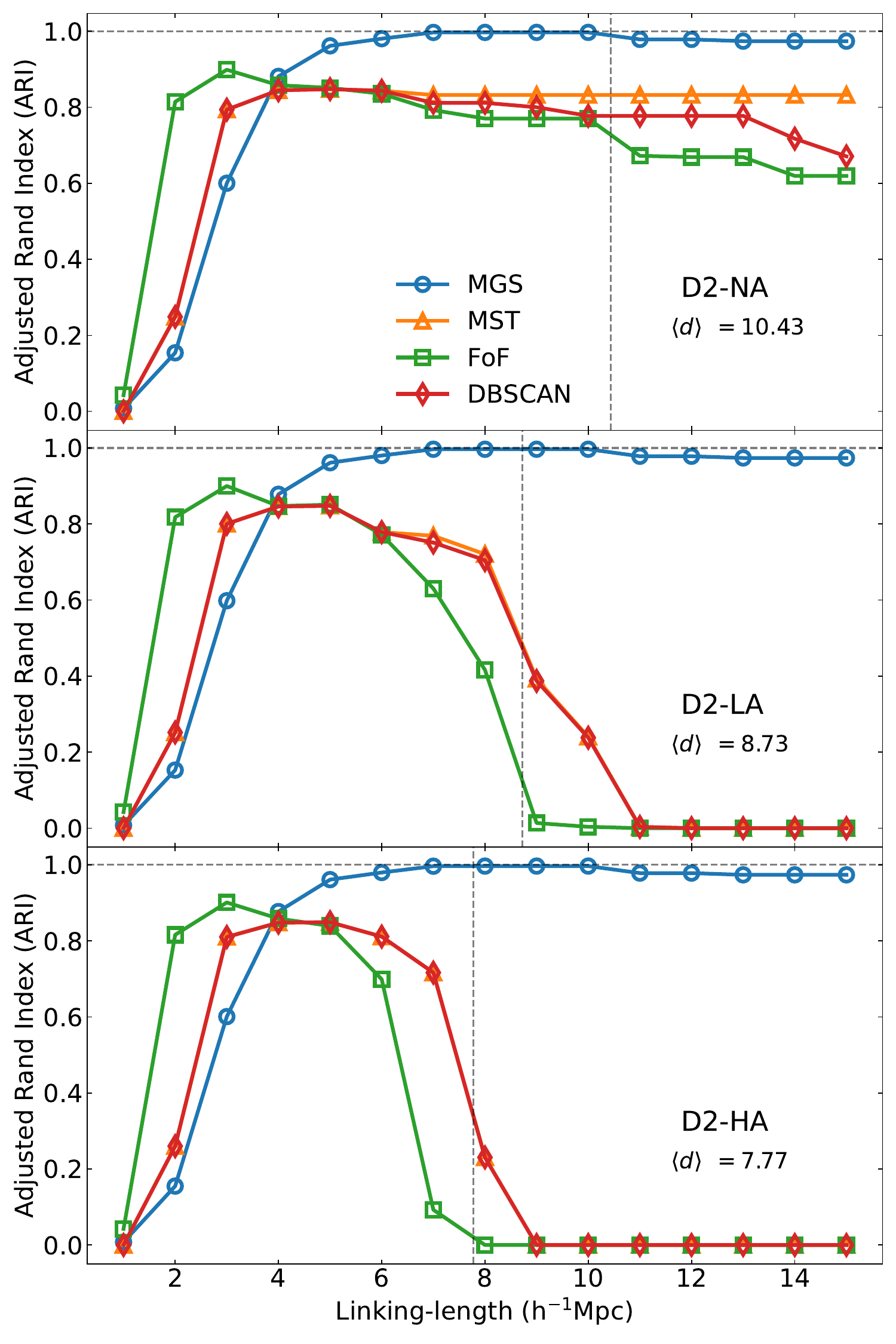}
\caption{Same as figure~\ref{fig:4alg1}, but with D2-NA(top), D2-LA(middle), and D2-HA(bottom).
The vertical dashed line is the mean separation of data ($\langle d \rangle$). 
 Since each data set has a different overall number density.}
\label{fig:4alg2}
\end{figure}

Figure~\ref{fig:4alg2} shows the results for controlled data set 2, which includes the D2-NA data with no additional galaxies and the D2-LA and D2-HA data with additional galaxies. The top panel shows the ARI identified by each algorithm for the D2-NA data. When this data was generated, the minimum number of galaxies per network was set to 50. In the region of small linking lengths, all algorithms identify fewer than 50 networks and obtain low ARI because the linking length is too small to connect the galaxies in the networks. 
As a result, the networks identified by the algorithms have fewer than 50 member galaxies and are therefore not considered as true networks. 
For larger linking lengths, particularly those larger than 5, the difference between the MGS algorithm and the other algorithms becomes more pronounced. The MGS algorithm continues to accurately identify the correct number of networks, while the other algorithms identify fewer networks due to the merging of originally separate networks.

The behavior of the algorithms with additional galaxies is even more distinct. The middle panel of figure~\ref{fig:4alg2} shows the results for the D2-LA data, where the other three algorithms identify only a single network for very large linking lengths. As the linking length increases, the algorithms merge several networks into a single giant network, resulting in a significantly lower number of networks than the original data and consequently a low ARI value. 
This rapid increment of a single giant network is called ``percolation,'' and it is known to occur at linking length similar to the mean-separation ($\ell \simeq \langle d \rangle$) for the ideal random Poisson graph \citep{Dall2002}.
Figure~\ref{fig:4alg2} clearly shows that such percolation occurs at $\ell \simeq \langle d \rangle$ for all three benchmark algorithms.

\begin{figure}[t]
\centering
\includegraphics[width=0.8\textwidth]{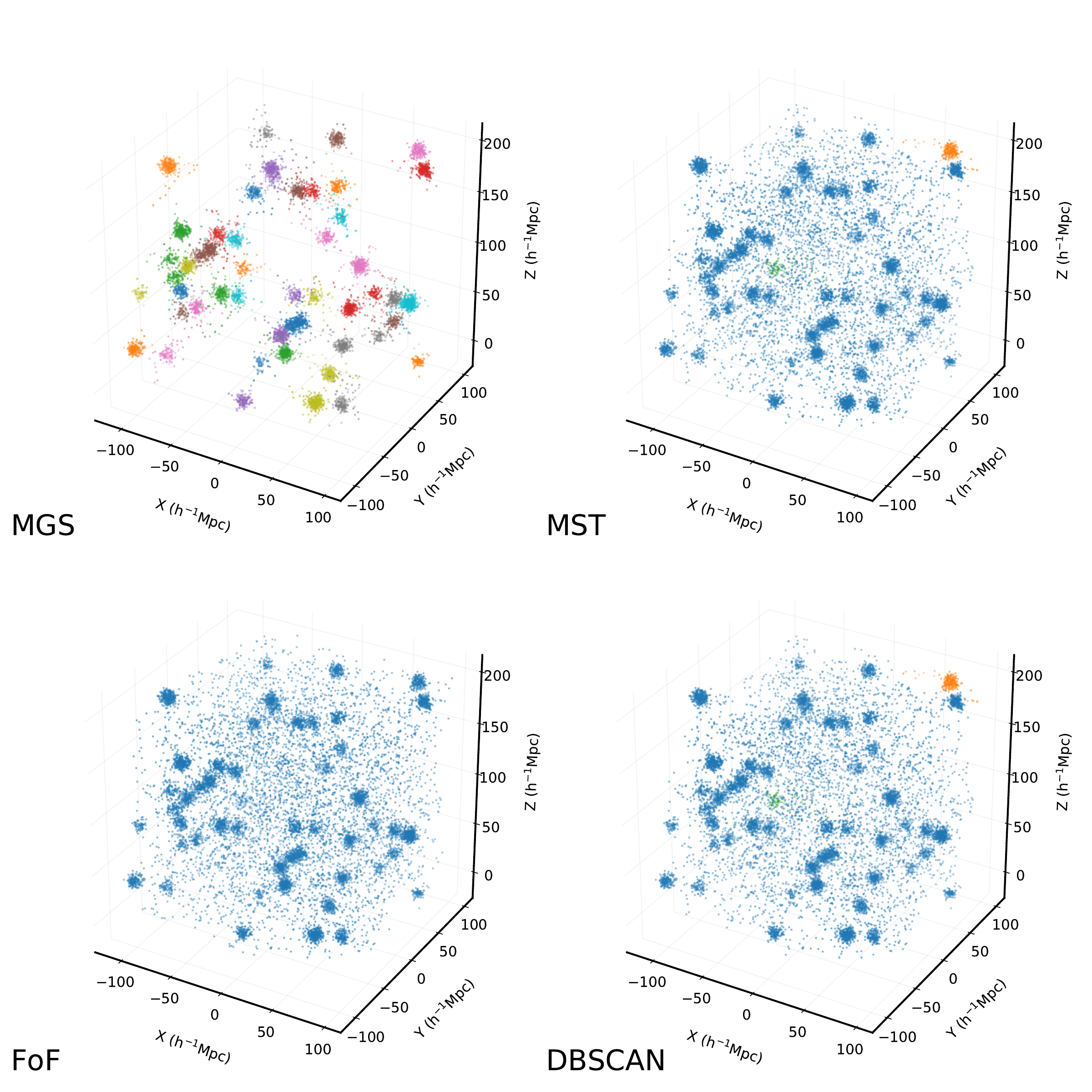}
\caption{3D distribution of clustering results from MGS and other benchmark algorithms in D2-LA with linking length $\ell = 11\Mpch$, which is longer than the mean-separation $\langle d \rangle = 8.73\Mpch$.
Color indicates the network membership.
MGS finds 49 networks among 50 true networks, while other algorithms connect most of the galaxies and finally make a giant network (blue color).}
\label{fig:2alg}
\end{figure}

Note that, however, FoF has the shortest linking length at percolation, while both MST and DBSCAN share a similar value of linking length at percolation.
This is because FoF does not have an additional consideration for limiting the network boundary that exists in the other two algorithms (minimizing the number of edges in MST and core definition in DBSCAN).
Figure~\ref{fig:2alg} shows the 3D distributions of clustering results from various algorithms at linking length $\ell = 11\Mpch$, which is longer than the mean separation $\langle d \rangle = 8.73 \Mpch$.
As expected, three benchmark algorithms show percolation (blue color), while our MGS algorithm successfully reconstructs most of the true networks.
Note that only one giant network is found in the FoF algorithm, while both MST and DBSCAN have two additional small networks (green and orange colors).

The behavior of the MGS algorithm for the D2-HA data is slightly different. In the region of small linking lengths, the MGS algorithm accurately identifies the correct number of networks. 
However, for larger linking lengths, particularly $\ell > 13\Mpch$, the MGS algorithm identifies additional networks that were not present in the original data (corresponding to a low ARI value). These ``fake'' networks are not true networks and are not representative of the underlying structure of the data. This behavior highlights the ability of the MGS algorithm to identify networks in data with a complex distribution of galaxies but also underscores the importance of choosing an appropriate linking length to avoid identifying false networks.

\begin{figure}[t]
\centering
\includegraphics[width=0.6\textwidth]{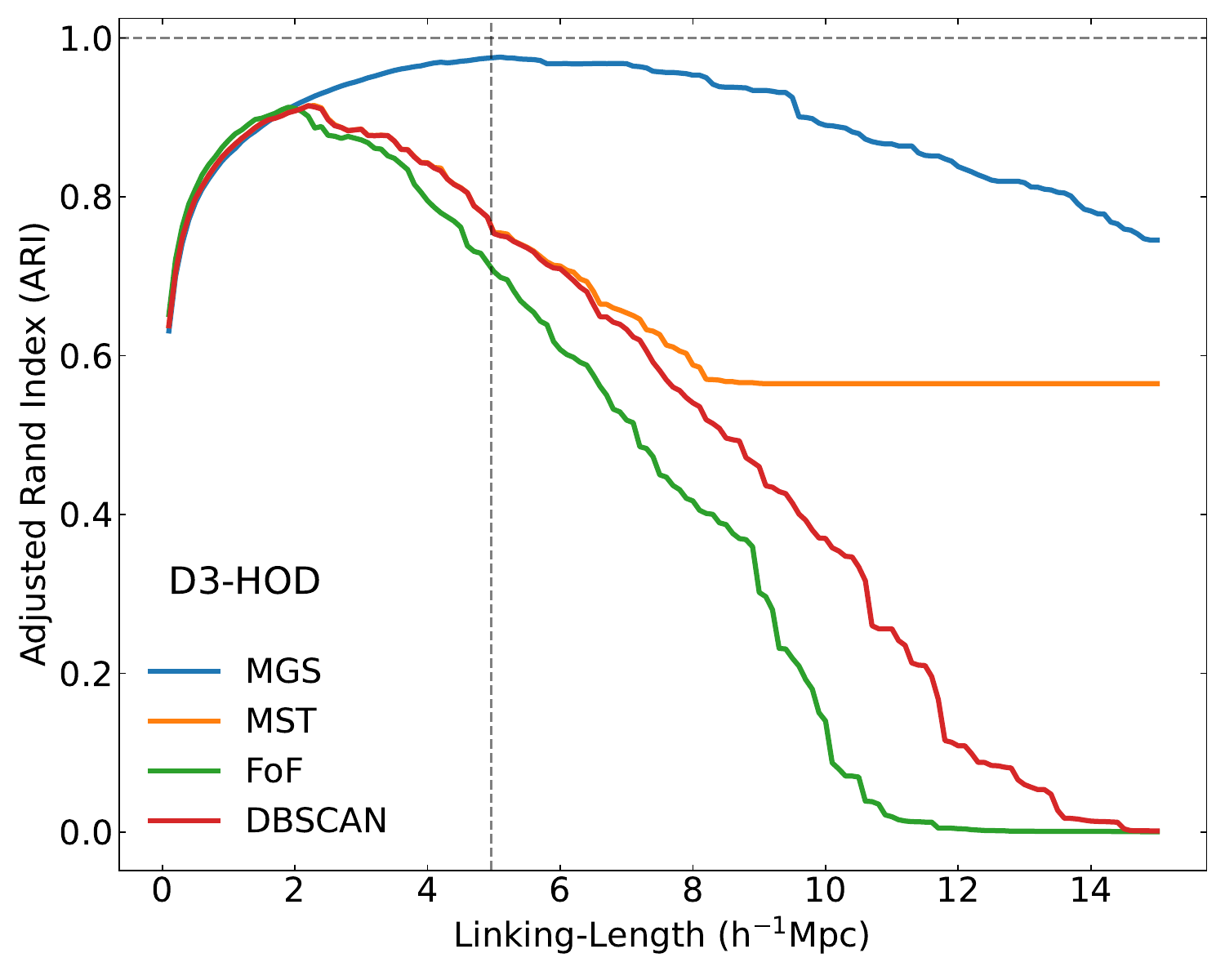}
\caption{Same as figures~\ref{fig:4alg1}--\ref{fig:4alg2}, but with D3-HOD.}
\label{fig:4alg3}
\end{figure}

Figure~\ref{fig:4alg3} shows the number of networks for controlled 3 data with a more complex environment than D1--D2. 
At $\ell \gtrsim \langle d \rangle /2 \approx 3 \Mpch$, the number of networks and the ARI value using FoF and DBSCAN decreases as the linking length increases, resulting in the percolation at $\ell \gtrsim \langle d \rangle$. The MST shows a flat curve when the linking length is larger than $\sim 8\Mpch$.
This is because MST connects all galaxies with minimal edge first, and then we cut off the links with linking length. 
Therefore, if there were no links longer than $8\Mpch$ in the original tree, then cutting the links with any longer linking length than $8\Mpch$ would not change the result.
So, the ARI value using MST shows a constant value. 

In contrast, the ARI value and also the number of networks identified by the MGS algorithm slowly decrease as the linking length increases. This is because the networks in this data set are close to each other and are easily merged by the algorithm for large linking lengths. However, the MGS algorithm is able to identify networks based on density, which allows it to retain the structure of the networks even for large linking lengths. This is the main advantage of the MGS algorithm compared to the other three algorithms, which are not able to accurately identify networks in complex data sets.

%- - - - - - - - - - - - - - - - - - - - - - - - - - - - - - - - - - - - - - - -
\subsection{Results with Observational and Cosmological Simulation Data}\label{sec:results_real}
%- - - - - - - - - - - - - - - - - - - - - - - - - - - - - - - - - - - - - - - -

\begin{figure}[t]
\centering
\includegraphics[width=0.8\textwidth]{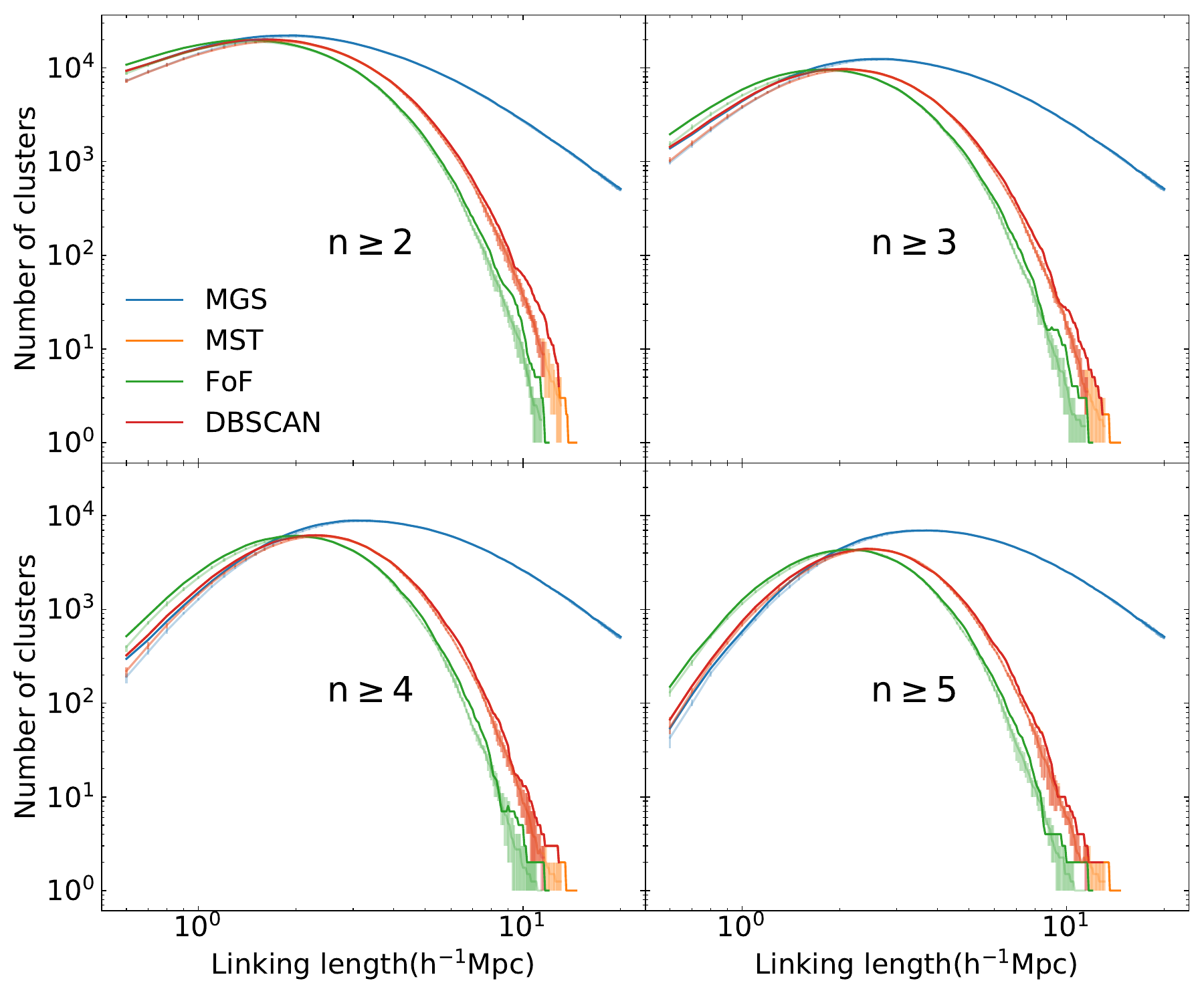}
\caption{The number of networks as a function of linking length for D4-SDSS (thick lines) and D4-HR4.
For D4-HR4, the average values and the ranges between the minimums and the maximums of 4 data samples are drawn as thin lines and error bars.
Each panel shows the results from the different choices of the minimum number of member galaxies to identify networks.
}
\label{fig:4alg4}
\end{figure}

Figure~\ref{fig:4alg4} shows the results of the four algorithms applied to both KIAS-VAGC observational data and four sets of HR4 lightcone data. In the previous subsections, we calculated the ARI in order to evaluate the algorithms. One of the reasons why we used the ARI is that it can show the performance of the clustering qualitatively since we already have the original information of the network. 
However, with observational data, we do not know the true networks.
Consequently, we can only track the number of detected networks changing with both linking lengths and with the minimum number of member galaxies from 2 to 5, and comment on the results.

For all four clustering algorithms, the HR4 simulation results match well with the observations within cosmic variance, especially for $n \geq 5$ at $\ell \gtrsim \langle d_{\rm particle} \rangle = 0.5 \Mpch$.
On the other hand, HR4 tends to underestimate the number of networks for a smaller minimum number of member galaxies and/or smaller linking length $\ell \lesssim 0.5 \Mpch$.
This may mean that, despite the agreement with the observation in terms of 2pCF below $1\Mpch$-scale, some disagreements exist between HR4 and observation in terms of the higher-order statistics in smaller scales than the particle mean separation scale.

One notable feature of the MGS algorithm is that it does not create a single giant network for large linking lengths. Instead, the algorithm identifies a number of smaller networks, even for large linking lengths. This is in contrast to the other three algorithms, which all create a single giant network for large linking lengths. This difference highlights the ability of the MGS algorithm to accurately identify networks in data with a complex distribution of galaxies.

\begin{figure}[t]
\centering
\includegraphics[width=0.8\textwidth]{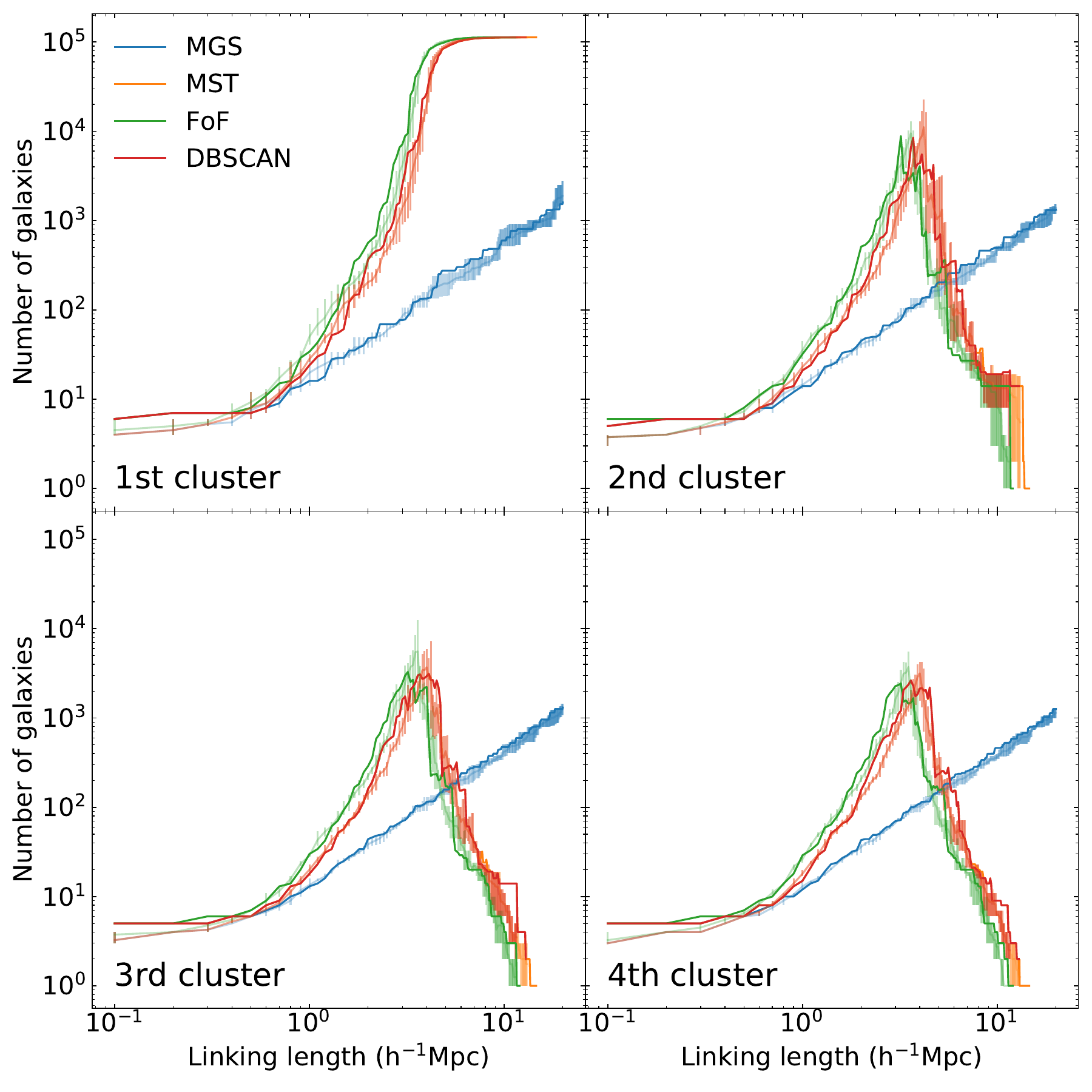}
\caption{The number of member galaxies for the 1st, 2nd, 3rd and 4th richest networks in D4-SDSS (thick lines) and  in D4-HR4 as a function of linking length.
For D4-HR4, the average values and the ranges between the minimums and the maximums of 4 data samples are drawn as thin lines and error bars.
Color indicates the different clustering algorithms.}
\label{fig:4alg5}
\end{figure}

Figure~\ref{fig:4alg5} shows the number of member galaxies for the 1st to 4th richest networks identified by each algorithm in D4-SDSS and D4-HR4.
Similar to figure~\ref{fig:4alg4}, both results from the simulation and observation data match well with each other.
The top left panel of the figure shows the shape of the richest network for each algorithm. 
As the linking length increases over a certain value, the richest network identified by the FoF, MST, and DBSCAN algorithms contains all of the galaxies in the data, while the MGS algorithm identifies a network with only a portion of the galaxies. 
This indicates that the MGS algorithm is able to identify multiple networks even for large linking lengths, while the other algorithms merge all of the galaxies into a single giant network.

The main difference between the MGS algorithm and the other three algorithms becomes particularly clear when examining the number of member galaxies in the 2nd to 4th richest networks (upper right and bottom panels of figure~\ref{fig:4alg5}).
As the linking length increases, the number of member galaxies in these networks identified by the FoF, MST, and DBSCAN algorithms decreases to zero. 
This is because the first richest network identified by these algorithms took all the galaxies in the data, leaving no galaxies to be considered for further clustering.
In contrast, the MGS algorithm is able to identify multiple networks even with large linking lengths as the richest network does not monopolize all galaxies. 
This demonstrates the ability of the MGS algorithm to accurately identify networks in data with a complex distribution of galaxies.

\subsection{Application to cosmology}\label{sec:results_cosmo}
%We have discussed the MGS based on our controlled and observational data. These results illustrate the algorithm's mechanism and its capability to determine a more stable network structure. We further investigate the applicability of our algorithm in cosmology, focusing on the %halo 
%{\color{OliveGreen} galaxy cluster} mass function, a useful method for identifying structures in the universe~\cite{peebles1993, peacock1999}. %The halo mass function reveals the number of halos within a given mass range, with various theoretical predictions and their definitions~\cite{Press1974, Sheth1999, Jenkins2001}. Through the halo mass function, we assess the reliability of halos identified by our algorithm within the framework of the standard cosmology.
%
%As detailed in section~\ref{sec:data_real_hr4}, the {\color{OliveGreen} HR4} simulation provides proxy masses for mock galaxies. %Utilizing this mass information,
%{\color{OliveGreen} Although the exact definition of galaxy ``mass'' here is rather complicated (see ref.~\cite{hong2016} for details),} we {\color{OliveGreen} can use it to roughly} estimate the %halo
%{\color{OliveGreen} galaxy cluster} mass function. %within a specified mass range.

We have discussed the MGS based on our controlled and observational data. These results illustrate the algorithm's mechanism and its capability to determine a more stable network structure. We further investigate the applicability of our algorithm in cosmology, focusing on the galaxy cluster mass function, a useful method for identifying structures in the universe~\cite{peebles1993, peacock1999}. As detailed in section~\ref{sec:data_real_hr4}, the HR4 simulation provides proxy masses for mock galaxies. Although the exact definition of galaxy ``mass'' here is rather complicated (see ref.~\cite{hong2016} for details), we can use it to roughly estimate the galaxy cluster mass function.

To generate clusters from galaxies, it is essential to set an appropriate linking length for the algorithms. The conventional way for determining the linking length involves using the linking parameter, $b$, where the linking length is defined as $b$ times the mean separation. In cosmological studies, $b$ is often set to $0.2$ for the FoF algorithm (e.g., see refs.~\cite{Robotham2011, Yang2021}, while the selection of the linking parameter varies based on the specific objectives of individual researches~\cite{Duarte2014}), but comparing the same linking parameter for different algorithms is uneven due to their distinct properties in finding clusters. Therefore, we varied $b$ from $0.1$ to $0.4$ to examine the variation in 
galaxy cluster mass functions for both FoF and MGS with the mass range $2\times10^{12} - 2\times10^{15} \Msunh$. 

% The results are presented in figure~\ref{fig:massfunc}, where we averaged the %halo 
% {\color{OliveGreen} galaxy cluster} mass function for each mass bin from four {\color{OliveGreen} D4-HR4} simulations. Overall, the %halo 
% {\color{OliveGreen} shapes of } mass function %shapes 
% from both algorithms exhibit a reliable power-law feature, as predicted by theory. At low mass, %the halo 
% {\color{OliveGreen} both} mass function{\color{OliveGreen}s} deviate %s 
% from the power-law due to the limitation on the number of %halo 
% {\color{OliveGreen} cluster} members and the low mass range. At high mass, MGS and FoF display different trends. MGS shows %stable halo mass functions with various 
% {\color{OliveGreen} a similar shape of mass function by changing} linking parameters, while FoF exhibits a large transition. %This behavior is 
% {\color{OliveGreen} As already} discussed in section~%\ref{sec:results}
% \ref{sec:results_real}{\color{OliveGreen}, this is mainly because }%. As the linking parameter increases, 
% FoF forms one giant network encompassing smaller networks {\color{OliveGreen} at high linking parameter}. Consequently, the FoF %halo 
% {\color{OliveGreen} galaxy cluster} mass function %shows much higher mass of halos 
% {\color{OliveGreen} has more massive clusters with mass greater than $\sim 10^{14} \Msunh$} than MGS.

\begin{figure}
\centering
\includegraphics[width=0.8\textwidth]{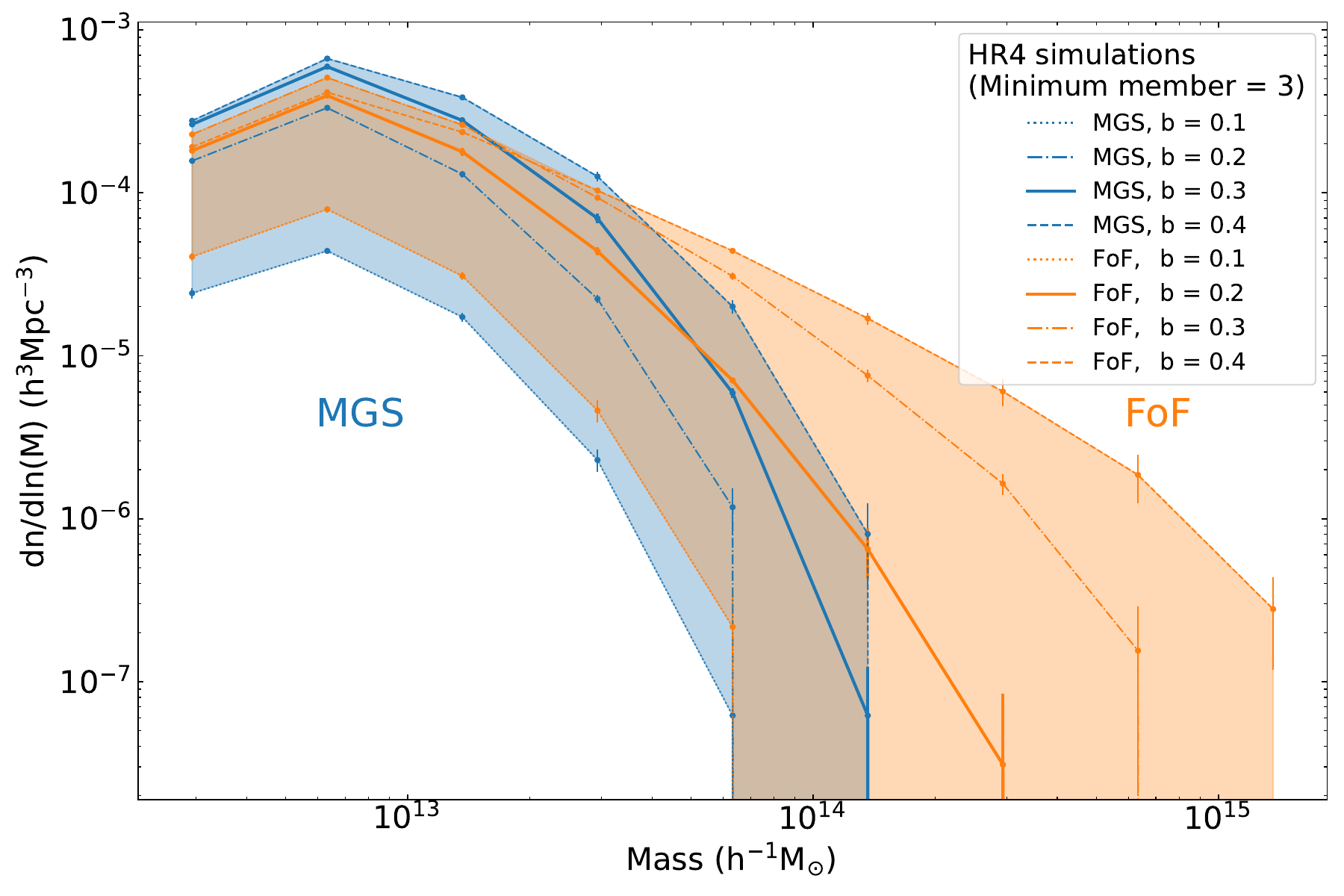}
\caption{
The averaged mass functions of galaxy clusters having at least three member galaxies found by the MGS (blue) and FoF (orange) algorithms by varying linking parameters ($b = 0.1-0.4$; line styles), derived from four D4-HR4 simulations. The error bars represent one standard deviation obtained from four simulation results. The colored shaded regions denote the ranges of mass functions from a given algorithm by varying the linking parameter.}
\label{fig:massfunc}
\end{figure}

The results are presented in figure~\ref{fig:massfunc}, where we averaged the 
galaxy cluster mass function for each mass bin from four D4-HR4 simulations. Overall, the shapes of mass function from both algorithms exhibit a reliable power-law feature, as predicted by theory. At low mass, both mass functions deviate from the power-law due to the limitation on the number of cluster members and the low mass range. At high mass, MGS and FoF display different trends. MGS shows a similar shape of mass function by changing linking parameters, while FoF exhibits a large transition. As already discussed in section~\ref{sec:results_real}, this is mainly because FoF forms one giant network encompassing smaller networks at high linking parameter. Consequently, the FoF galaxy cluster mass function has more massive clusters with mass greater than $\sim 10^{14} \Msunh$ than MGS.

% It is noteworthy that the shape of the halo mass function with %$b$ set to $0.2$ 
% {\color{OliveGreen} $b = 0.2$} for FoF is similar to %$b$ set to $0.3$
% {\color{OliveGreen} $b = 0.3$} for MGS (thick lines in figure~\ref{fig:massfunc}). This demonstrates that the appropriate linking length (or, linking parameter) differs between algorithms. {\color{OliveGreen} Interestingly, the ratio between two choices of linking lengths ($\ell_{\rm MGS} / \ell_{\rm FoF} = b_{\rm MGS} / b_{\rm FoF} \simeq 1.5$) is similar to such ratio for D3-HOD with the choices of best-ARI (figure~\ref{fig:4alg3}), as well as the ratio for D4-HR4 with the choices of maximizing the number of networks (figure~\ref{fig:4alg4}) for each clustering algorithm.} 

It is noteworthy that the shape of the halo mass function with $b = 0.2$ for FoF is similar to $b = 0.3$ for MGS (thick lines in figure~\ref{fig:massfunc}). This demonstrates that the appropriate linking length (or, linking parameter) differs between algorithms. Interestingly, the ratio between two choices of linking lengths ($\ell_{\rm MGS} / \ell_{\rm FoF} = b_{\rm MGS} / b_{\rm FoF} \simeq 1.5$) is similar to such ratio for D3-HOD with the choices of best-ARI (figure~\ref{fig:4alg3}), as well as the ratio for D4-HR4 with the choices of maximizing the number of networks (figure~\ref{fig:4alg4}) for each clustering algorithm.

% To further investigate this difference in distribution, we examine the structure of %halos 
% {\color{OliveGreen} galaxy clusters} constructed by FoF and MGS. The 3D distributions of MGS and FoF {\color{OliveGreen} clusters at a certain identical region from one D4-HR4 simulation} are presented in figure~\ref{fig:mgsfof}%. This case provides a well-contrasted example of FoF and MGS clusters
% , {\color{blue} allowing us to specifically illustrate the relationship between halos and clusters.} 
% {\color{magenta} In the left panel the large FoF cluster comprises a very elongated collection of galaxies. The most massive galaxy resides near the upper focus of the ellipsoid where there is a clear over-density, while the lower focus contains another collection of galaxies. In the right-hand panel, however, MGS has separated this collection of galaxies into three distinct groups, with the most massive galaxy in one upper grouping and the lower collection of galaxies being separated into their own two groups.} 
% %It represents that separated distributions of galaxies within one giant FoF cluster at the bottom right, while MGS divides it into three independent clusters. Regarding the mass distribution of galaxies, the heaviest galaxy is located within the largest cluster of MGS. 
% These behaviors align with the implications of the %halo 
% {\color{OliveGreen} cluster} mass function, which exhibits a long tail at high mass due to FoF connecting adjacent clusters into one giant cluster.

\begin{figure*}
\centering
\includegraphics[width=1.0\textwidth]{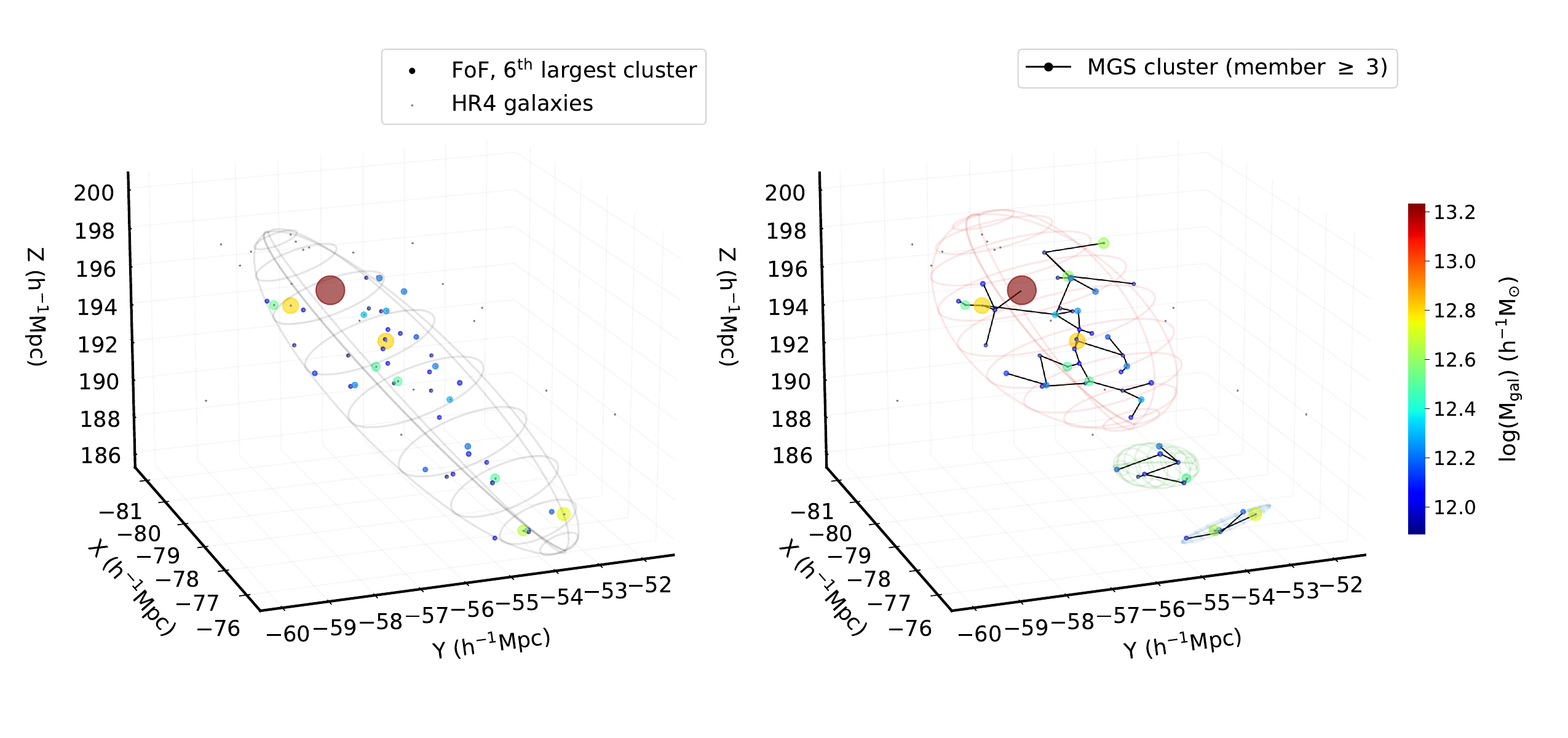}
\caption{The 3D distribution of member galaxies (circles) in the 6th largest FoF cluster from one D4-HR4 data (left) and the corresponding MGS clusters (right). Color and size of circles show the galaxy ``mass'' (see ref.~\cite{hong2016} for details). Gray dots are non-member galaxies. Ellipsoid fitting is applied to the galaxies in an identical cluster. In the case of MGS, connections between galaxies are drawn using network information, indicated by black lines.}
\label{fig:mgsfof}
\end{figure*}

To further investigate this difference in distribution, we examine the structure of galaxy clusters constructed by FoF and MGS. The 3D distributions of MGS and FoF clusters at a certain identical region from one D4-HR4 simulation are presented in figure~\ref{fig:mgsfof}, allowing us to specifically illustrate the relationship between halos and clusters. In the left panel, the large FoF cluster comprises a very elongated collection of galaxies. The most massive galaxy resides near the upper focus of the ellipsoid where there is a clear over-density, while the lower focus contains another collection of galaxies. In the right-hand panel, however, MGS has separated this collection of galaxies into three distinct groups, with the most massive galaxy in one upper grouping and the lower collection of galaxies being separated into their own two groups. These behaviors align with the implications of the cluster mass function, which exhibits a long tail at high mass due to FoF connecting adjacent clusters into one giant cluster.

% \begin{figure*}
% \centering
% % \includegraphics[width=0.8\textwidth]
% \includegraphics[width=1.0\textwidth]{fig_3D_dist_mgs_fof.pdf}
% \caption{The 3D distribution %is presented for 
% {\color{OliveGreen} of member galaxies (circles) in} the 6th largest FoF cluster {\color{blue} from %among the 
% one D4-HR4 data} (left) and the corresponding MGS clusters (right). %Member galaxies of the cluster are denoted as circles, while isolated galaxies are represented as points. The color of each circle signifies the mass of the corresponding galaxy, and the size of the circle is proportional to the galaxy's mass, scaled accordingly.
% {\color{OliveGreen} Color and size of circles show the galaxy ``mass'' (see ref.~\cite{hong2016} for details). Gray dots are non-member galaxies.} Ellipsoid fitting is applied to the galaxies {\color{OliveGreen} in an identical cluster.} %, and the color of the ellipsoids is arbitrary. 
% In the case of MGS, connections between galaxies are drawn using network information, indicated by black lines.}
% \label{fig:mgsfof}
% \end{figure*}

\section{Conclusions}\label{sec:conclusions}

The MulGuisin (MGS) algorithm is a powerful technique for identifying networks in data from astrophysical simulations and observations. It consistently produces results closer to those inferred from human visual inspection. In comparison to other clustering algorithms, such as the friends-of-friends (FoF) algorithm, the minimum spanning tree (MST) algorithm, and the DBSCAN algorithm, the MGS algorithm has several advantages. The MGS algorithm is able to take into consideration the local density and is able to accurately identify networks even in complex data sets with a large number of galaxies. In contrast, the FoF, MST, and DBSCAN algorithms often merge networks into a single giant network for large linking lengths, losing the ability to accurately identify individual networks.
This characteristic of the MGS algorithm is particularly important for analyzing data from astrophysical simulations and observations.

In this proof of concept work, we have shown that the jet-finding algorithm MGS can be applied to mock galaxy data resulting in reliable network identification. However, the identification of networks in real observation is a difficult issue due to survey incompleteness, selection effects, redshift-space distortions, etc. In future work, we will test MGS in the presence of realistic observational systematic effects. 

MGS also provides auxiliary topological information such as the number and length of connections for each galaxy. In future work, we will explore the use of this enhanced information in testing or constraining cosmological models.

\acknowledgments

The authors thank Changbom Park, Dongsu Bak, Ena Choi, and Jubee Sohn for helpful discussions.

This research was supported by the Basic Science Research Program through the National Research Foundation of Korea(NRF) funded by the Ministry of Education (2018\-R1\-A6\-A1\-A06024977) for Y.J..
C.G.S. was supported via the Basic Science Research Program from the National Research Foundation of South Korea (NRF) funded by the Ministry of Education (2020\-R1\-I1\-A1\-A01073494).
S.E.H. was supported by the projects \begin{CJK}{UTF8}{mj}우주거대구조를 이용한 암흑우주 연구\end{CJK} 
(``Understanding Dark Universe Using Large Scale Structure of the Universe'') and \begin{CJK}{UTF8}{mj}가속팽창하는 우주의 원리에 관한 연구\end{CJK}(``Research on the Principle of Accelerated Expansion of the Universe''), funded by the Ministry of Science.
This work was supported by the 2023 Research Fund of the University of Seoul for I.P..

This work was supported by the Supercomputing Center/Korea Institute of Science and Technology Information, with supercomputing resources including technical support (KSC-2013-G2-003), and the simulation data were transferred through a high-speed network provided by KREONET/GLORIAD.

We would like to acknowledge the Korean Astronomy and Machine Learning (KAML) meeting series for providing a forum for fruitful discussions on the intersection of astronomy and machine learning.

Funding for the SDSS and SDSS-II has been provided by the Alfred P. Sloan Foundation, the Participating Institutions, the National Science Foundation, the US Department of Energy,
the National Aeronautics and Space Administration, the Japanese Monbukagakusho, the Max Planck Society, and the Higher Education Funding Council for England. 
The SDSS website is \url{http://www.sdss.org/}.

The SDSS is managed by the Astrophysical Research Consortium for the Participating Institutions. The Participating Institutions are the American Museum of Natural History,
Astrophysical Institute Potsdam, University of Basel, University of Cambridge, Case Western Reserve University, University of Chicago, Drexel University, Fermilab, the
Institute for Advanced Study, the Japan Participation Group, Johns Hopkins University, the Joint Institute for Nuclear Astrophysics, the Kavli Institute for Particle Astrophysics and Cosmology, the Korean Scientist Group, the Chinese Academy of Sciences (LAMOST), Los Alamos National Laboratory, Max Planck Institute for Astronomy (MPIA), the Max Planck
Institute for Astrophysics (MPA), New Mexico State University, Ohio State University, University of Pittsburgh, University of Portsmouth, Princeton University, the US Naval
Observatory, and the University of Washington.

\paragraph{Data Availability}

The up-to-date MulGuisin algorithm can be downloaded at \url{https://github.com/youngju20/Mulguisin}.

%===============================================================================
\bibliographystyle{JHEP}
\bibliography{main}

\providecommand{\href}[2]{#2}\begingroup\raggedright\begin{thebibliography}{10}

\bibitem{White1978}
S.D.M.~{White} and M.J.~{Rees}, \emph{{Core condensation in heavy halos: a
  two-stage theory for galaxy formation and clustering.}},
  \href{https://doi.org/10.1093/mnras/183.3.341}{\emph{\mnras} {\bfseries 183}
  (1978) 341}.

\bibitem{Fall1980}
S.M.~{Fall} and G.~{Efstathiou}, \emph{{Formation and rotation of disc galaxies
  with haloes.}}, \href{https://doi.org/10.1093/mnras/193.2.189}{\emph{\mnras}
  {\bfseries 193} (1980) 189}.

\bibitem{Blumenthal1984}
G.R.~{Blumenthal}, S.M.~{Faber}, J.R.~{Primack} and M.J.~{Rees},
  \emph{{Formation of galaxies and large-scale structure with cold dark
  matter.}}, \href{https://doi.org/10.1038/311517a0}{\emph{\nat} {\bfseries
  311} (1984) 517}.

\bibitem{Gott1986}
I.~{Gott}, J.~Richard, A.L.~{Melott} and M.~{Dickinson}, \emph{{The Sponge-like
  Topology of Large-Scale Structure in the Universe}},
  \href{https://doi.org/10.1086/164347}{\emph{\apj} {\bfseries 306} (1986)
  341}.

\bibitem{Park1991}
C.~{Park} and I.~{Gott}, J.~R., \emph{{Dynamical Evolution of Topology of
  Large-Scale Structure}}, \href{https://doi.org/10.1086/170445}{\emph{\apj}
  {\bfseries 378} (1991) 457}.

\bibitem{Park2010}
C.~{Park} and Y.-R.~{Kim}, \emph{{Large-scale Structure of the Universe as a
  Cosmic Standard Ruler}},
  \href{https://doi.org/10.1088/2041-8205/715/2/L185}{\emph{\apjl} {\bfseries
  715} (2010) L185} [\href{https://arxiv.org/abs/0905.2268}{{\ttfamily
  0905.2268}}].

\bibitem{Appleby2017}
S.~{Appleby}, C.~{Park}, S.E.~{Hong} and J.~{Kim}, \emph{{Topology of
  Large-Scale Structures of Galaxies in two Dimensions{\textemdash}Systematic
  Effects}}, \href{https://doi.org/10.3847/1538-4357/836/1/45}{\emph{\apj}
  {\bfseries 836} (2017) 45}
  [\href{https://arxiv.org/abs/1702.04511}{{\ttfamily 1702.04511}}].

\bibitem{Appleby2018}
S.~{Appleby}, P.~{Chingangbam}, C.~{Park}, S.E.~{Hong}, J.~{Kim} and
  V.~{Ganesan}, \emph{{Minkowski Tensors in Two Dimensions: Probing the
  Morphology and Isotropy of the Matter and Galaxy Density Fields}},
  \href{https://doi.org/10.3847/1538-4357/aabb53}{\emph{\apj} {\bfseries 858}
  (2018) 87} [\href{https://arxiv.org/abs/1712.07466}{{\ttfamily 1712.07466}}].

\bibitem{Alcock1979}
C.~{Alcock} and B.~{Paczynski}, \emph{{An evolution free test for non-zero
  cosmological constant}}, \href{https://doi.org/10.1038/281358a0}{\emph{\nat}
  {\bfseries 281} (1979) 358}.

\bibitem{Ballinger1996}
W.E.~{Ballinger}, J.A.~{Peacock} and A.F.~{Heavens}, \emph{{Measuring the
  cosmological constant with redshift surveys}},
  \href{https://doi.org/10.1093/mnras/282.3.877}{\emph{\mnras} {\bfseries 282}
  (1996) 877} [\href{https://arxiv.org/abs/astro-ph/9605017}{{\ttfamily
  astro-ph/9605017}}].

\bibitem{Li2014}
X.-D.~{Li}, C.~{Park}, J.E.~{Forero-Romero} and J.~{Kim}, \emph{{Cosmological
  Constraints from the Redshift Dependence of the Alcock-Paczynski Test: Galaxy
  Density Gradient Field}},
  \href{https://doi.org/10.1088/0004-637X/796/2/137}{\emph{\apj} {\bfseries
  796} (2014) 137} [\href{https://arxiv.org/abs/1412.3564}{{\ttfamily
  1412.3564}}].

\bibitem{hpark2019}
H.~{Park}, C.~{Park}, C.G.~{Sabiu}, X.-d.~{Li}, S.E.~{Hong}, J.~{Kim} et~al.,
  \emph{{Alcock-Paczynski Test with the Evolution of Redshift-space Galaxy
  Clustering Anisotropy}},
  \href{https://doi.org/10.3847/1538-4357/ab2da1}{\emph{\apj} {\bfseries 881}
  (2019) 146} [\href{https://arxiv.org/abs/1904.05503}{{\ttfamily
  1904.05503}}].

\bibitem{Sheth1996}
R.K.~{Sheth}, \emph{{The distribution of pairwise peculiar velocities in the
  non-linear regime}},
  \href{https://doi.org/10.1093/mnras/279.4.1310}{\emph{\mnras} {\bfseries 279}
  (1996) 1310} [\href{https://arxiv.org/abs/astro-ph/9511068}{{\ttfamily
  astro-ph/9511068}}].

\bibitem{DeRose2019}
J.~{DeRose}, R.H.~{Wechsler}, J.L.~{Tinker}, M.R.~{Becker}, Y.-Y.~{Mao},
  T.~{McClintock} et~al., \emph{{The AEMULUS Project. I. Numerical Simulations
  for Precision Cosmology}},
  \href{https://doi.org/10.3847/1538-4357/ab1085}{\emph{\apj} {\bfseries 875}
  (2019) 69} [\href{https://arxiv.org/abs/1804.05865}{{\ttfamily 1804.05865}}].

\bibitem{Tonegawa2020}
M.~{Tonegawa}, C.~{Park}, Y.~{Zheng}, H.~{Park}, S.E.~{Hong}, H.S.~{Hwang}
  et~al., \emph{{Cosmological Information from the Small-scale Redshift-space
  Distortion}}, \href{https://doi.org/10.3847/1538-4357/ab95ff}{\emph{\apj}
  {\bfseries 897} (2020) 17}
  [\href{https://arxiv.org/abs/2005.12159}{{\ttfamily 2005.12159}}].

\bibitem{Serra2013}
A.L.~{Serra} and A.~{Diaferio}, \emph{{Identification of Members in the Central
  and Outer Regions of Galaxy Clusters}},
  \href{https://doi.org/10.1088/0004-637X/768/2/116}{\emph{\apj} {\bfseries
  768} (2013) 116} [\href{https://arxiv.org/abs/1211.3669}{{\ttfamily
  1211.3669}}].

\bibitem{Gifford2013}
D.~{Gifford}, C.~{Miller} and N.~{Kern}, \emph{{A Systematic Analysis of
  Caustic Methods for Galaxy Cluster Masses}},
  \href{https://doi.org/10.1088/0004-637X/773/2/116}{\emph{\apj} {\bfseries
  773} (2013) 116} [\href{https://arxiv.org/abs/1307.0017}{{\ttfamily
  1307.0017}}].

\bibitem{Knebe2011}
A.~{Knebe}, S.R.~{Knollmann}, S.I.~{Muldrew}, F.R.~{Pearce},
  M.A.~{Aragon-Calvo}, Y.~{Ascasibar} et~al., \emph{{Haloes gone MAD: The
  Halo-Finder Comparison Project}},
  \href{https://doi.org/10.1111/j.1365-2966.2011.18858.x}{\emph{\mnras}
  {\bfseries 415} (2011) 2293}
  [\href{https://arxiv.org/abs/1104.0949}{{\ttfamily 1104.0949}}].

\bibitem{1985ApJ...292..371D}
M.~{Davis}, G.~{Efstathiou}, C.S.~{Frenk} and S.D.M.~{White}, \emph{{The
  evolution of large-scale structure in a universe dominated by cold dark
  matter}}, \href{https://doi.org/10.1086/163168}{\emph{\apj} {\bfseries 292}
  (1985) 371}.

\bibitem{boruvka1926jistem}
O.~Bor{\r{u}}vka, \emph{O jist{\'e}m probl{\'e}mu minim{\'a}ln{\'\i}m},
  {\emph{Pl\`{a}ce mor. p\v{r}\'{i}rodov\v{e} d. spol. v Brn\v{e} III (3)}
  (1926) 37}.

\bibitem{Yang2007}
X.~{Yang}, H.J.~{Mo}, F.C.~{van den Bosch}, A.~{Pasquali}, C.~{Li} and
  M.~{Barden}, \emph{{Galaxy Groups in the SDSS DR4. I. The Catalog and Basic
  Properties}}, \href{https://doi.org/10.1086/522027}{\emph{\apj} {\bfseries
  671} (2007) 153} [\href{https://arxiv.org/abs/0707.4640}{{\ttfamily
  0707.4640}}].

\bibitem{Rykoff2013}
E.S.~{Rykoff}, E.~{Rozo}, M.T.~{Busha}, C.E.~{Cunha}, A.~{Finoguenov},
  A.~{Evrard} et~al., \emph{{redMaPPer. I. Algorithm and SDSS DR8 Catalog}},
  \href{https://doi.org/10.1088/0004-637X/785/2/104}{\emph{\apj} {\bfseries
  785} (2014) 104} [\href{https://arxiv.org/abs/1303.3562}{{\ttfamily
  1303.3562}}].

\bibitem{ester1996density}
M.~Ester, H.-P.~Kriegel, J.~Sander, X.~Xu et~al., \emph{A density-based
  algorithm for discovering clusters in large spatial databases with noise.},
  in \emph{kdd}, vol.~96, pp.~226--231, 1996.

\bibitem{Bosman:1998jfl}
M.~{Bosman}, I.~{Park}, M.~{Corbal}, D.~{Costanzo}, S.~{Lami}, R.~{Paoletti}
  et~al., \emph{Jet finder library: version 1.0}, {\emph{ATLAS Software}
  {\bfseries 98} (1998) 038}.

\bibitem{Huchra1982}
J.P.~Huchra and M.J.~Geller, \emph{Groups of galaxies i. nearby groups},
  \href{https://doi.org/10.1086/160000}{\emph{The Astrophysical Journal}
  {\bfseries 257} (1982) 423}.

\bibitem{Tago2008}
E.~Tago, J.~Einasto, E.~Saar, E.~Tempel, M.~Einasto, J.~Vennik et~al.,
  \emph{Groups of galaxies in the sdss data release 5},
  \href{https://doi.org/10.1051/0004-6361:20078036}{\emph{Astronomy \&
  Astrophysics} {\bfseries 479} (2008) 927}.

\bibitem{Duarte2014}
M.~Duarte and G.A.~Mamon, \emph{{How well does the friends-of-friends algorithm
  recover group properties from galaxy catalogues limited in both distance and
  luminosity?}}, \href{https://doi.org/10.1093/mnras/stu378}{\emph{Monthly
  Notices of the Royal Astronomical Society} {\bfseries 440} (2014) 1763}
  [\href{https://arxiv.org/abs/1401.0662}{{\ttfamily 1401.0662}}].

\bibitem{Tempel2016}
E.~Tempel, R.~Kipper, A.~Tamm, M.~Gramann, M.~Einasto, T.~Sepp et~al.,
  \emph{Friends-of-friends galaxy group finder with membership refinement},
  \href{https://doi.org/10.1051/0004-6361/201527755}{\emph{Astronomy \&
  Astrophysics} {\bfseries 588} (2016) A14}.

\bibitem{More2011}
S.~{More}, A.V.~{Kravtsov}, N.~{Dalal} and S.~{Gottl{\"o}ber}, \emph{{The
  Overdensity and Masses of the Friends-of-friends Halos and Universality of
  Halo Mass Function}},
  \href{https://doi.org/10.1088/0067-0049/195/1/4}{\emph{\apjs} {\bfseries 195}
  (2011) 4} [\href{https://arxiv.org/abs/1103.0005}{{\ttfamily 1103.0005}}].

\bibitem{Hearin_2017}
A.P.~Hearin, D.~Campbell, E.~Tollerud, P.~Behroozi, B.~Diemer, N.J.~Goldbaum
  et~al., \emph{Forward modeling of large-scale structure: An open-source
  approach with halotools},
  \href{https://doi.org/10.3847/1538-3881/aa859f}{\emph{The Astronomical
  Journal} {\bfseries 154} (2017) 190}.

\bibitem{Barrow1985}
J.D.~Barrow, S.P.~Bhavsar and D.H.~Sonoda, \emph{Minimal spanning trees,
  filaments and galaxy clustering},
  \href{https://doi.org/10.1093/mnras/216.1.17}{\emph{Monthly Notices of the
  Royal Astronomical Society} {\bfseries 216} (1985) 17}.

\bibitem{Krzewina1996}
L.G.~Krzewina and W.C.~Saslaw, \emph{{Minimal spanning tree statistics for the
  analysis of large-scale structure}},
  \href{https://doi.org/10.1093/mnras/278.3.869}{\emph{Monthly Notices of the
  Royal Astronomical Society} {\bfseries 278} (1996) 869}.

\bibitem{Naidoo2020}
K.~Naidoo, L.~Whiteway, E.~Massara, D.~Gualdi, O.~Lahav, M.~Viel et~al.,
  \emph{Beyond two-point statistics: using the minimum spanning tree as a tool
  for cosmology}, \href{https://doi.org/10.1093/mnras/stz3075}{\emph{Monthly
  Notices of the Royal Astronomical Society} {\bfseries 491} (2020) 1709}.

\bibitem{Naidoo2019}
K.~Naidoo, \emph{Mistree: a python package for constructing and analysing
  minimum spanning trees},
  \href{https://doi.org/10.21105/joss.01721}{\emph{Journal of Open Source
  Software} {\bfseries 4} (2019) 1721}.

\bibitem{sander2017dbscan}
J.~Sander, M.~Ester et~al., \emph{Dbscan revisited, revisited}, {\emph{ACM
  Transactions on Database Systems} {\bfseries 42} (2017) 1}.

\bibitem{scikit-learn}
F.~Pedregosa, G.~Varoquaux, A.~Gramfort, V.~Michel, B.~Thirion, O.~Grisel
  et~al., \emph{Scikit-learn: Machine learning in {P}ython}, {\emph{Journal of
  Machine Learning Research} {\bfseries 12} (2011) 2825}.

\bibitem{Kravtsov2004}
A.V.~Kravtsov, A.A.~Berlind, R.H.~Wechsler, A.A.~Klypin, S.~Gottlober,
  B.~Allgood et~al., \emph{{The Dark Side of the Halo Occupation
  Distribution}}, \href{https://doi.org/10.1086/420959/FULLTEXT/}{\emph{The
  Astrophysical Journal} {\bfseries 609} (2004) 35}
  [\href{https://arxiv.org/abs/0308519}{{\ttfamily 0308519}}].

\bibitem{choi2010}
Y.-Y.~{Choi}, D.-H.~{Han} and S.S.~{Kim}, \emph{{Korea Institute for Advanced
  Study Value-Added Galaxy Catalog}},
  \href{https://doi.org/10.5303/JKAS.2010.43.6.191}{\emph{Journal of Korean
  Astronomical Society} {\bfseries 43} (2010) 191}.

\bibitem{kim2015}
J.~{Kim}, C.~{Park}, B.~{L'Huillier} and S.E.~{Hong}, \emph{{Horizon Run 4
  Simulation: Coupled Evolution of Galaxies and Large-Scale Structures of the
  Universe}}, \href{https://doi.org/10.5303/JKAS.2015.48.4.213}{\emph{Journal
  of Korean Astronomical Society} {\bfseries 48} (2015) 213}.

\bibitem{hong2016}
S.E.~{Hong}, C.~{Park} and J.~{Kim}, \emph{{The Most Bound Halo Particle-Galaxy
  Correspondence Model: Comparison between Models with Different Merger
  Timescales}}, \href{https://doi.org/10.3847/0004-637X/823/2/103}{\emph{\apj}
  {\bfseries 823} (2016) 103}.

\bibitem{Press1974}
W.H.~{Press} and P.~{Schechter}, \emph{{Formation of Galaxies and Clusters of
  Galaxies by Self-Similar Gravitational Condensation}},
  \href{https://doi.org/10.1086/152650}{\emph{\apj} {\bfseries 187} (1974)
  425}.

\bibitem{Planck2016}
{Planck Collaboration}, P.A.R.~{Ade}, N.~{Aghanim}, M.~{Arnaud}, M.~{Ashdown},
  J.~{Aumont} et~al., \emph{{Planck 2015 results. XIII. Cosmological
  parameters}}, \href{https://doi.org/10.1051/0004-6361/201525830}{\emph{\aap}
  {\bfseries 594} (2016) A13}
  [\href{https://arxiv.org/abs/1502.01589}{{\ttfamily 1502.01589}}].

\bibitem{Diemer_2018}
B.~Diemer, \emph{{COLOSSUS}: A python toolkit for cosmology, large-scale
  structure, and dark matter halos},
  \href{https://doi.org/10.3847/1538-4365/aaee8c}{\emph{The Astrophysical
  Journal Supplement Series} {\bfseries 239} (2018) 35}.

\bibitem{Berlind2002}
A.A.~Berlind and D.H.~Weinberg, \emph{{The Halo Occupation Distribution: Toward
  an Empirical Determination of the Relation between Galaxies and Mass}},
  \href{https://doi.org/10.1086/341469}{\emph{The Astrophysical Journal}
  {\bfseries 575} (2002) 587} [\href{https://arxiv.org/abs/0109001}{{\ttfamily
  0109001}}].

\bibitem{Navarro1996}
J.F.~{Navarro}, C.S.~{Frenk} and S.D.M.~{White}, \emph{{The Structure of Cold
  Dark Matter Halos}}, \href{https://doi.org/10.1086/177173}{\emph{\apj}
  {\bfseries 462} (1996) 563}
  [\href{https://arxiv.org/abs/astro-ph/9508025}{{\ttfamily
  astro-ph/9508025}}].

\bibitem{2013SSRv..177..155L}
M.~{Limousin}, A.~{Morandi}, M.~{Sereno}, M.~{Meneghetti}, S.~{Ettori},
  M.~{Bartelmann} et~al., \emph{{The Three-Dimensional Shapes of Galaxy
  Clusters}}, \href{https://doi.org/10.1007/s11214-013-9980-y}{\emph{\ssr}
  {\bfseries 177} (2013) 155}
  [\href{https://arxiv.org/abs/1210.3067}{{\ttfamily 1210.3067}}].

\bibitem{Farrens:2011eb}
S.~Farrens, F.B.~Abdalla, E.S.~Cypriano, C.~Sabiu and C.~Blake,
  \emph{{Friends-of-Friends Groups and Clusters in the 2SLAQ Catalogue}},
  \href{https://doi.org/10.1111/j.1365-2966.2011.19356.x}{\emph{Mon. Not. Roy.
  Astron. Soc.} {\bfseries 417} (2011) 1402}
  [\href{https://arxiv.org/abs/1106.5687}{{\ttfamily 1106.5687}}].

\bibitem{2005AJ....129.2562B}
M.R.~{Blanton}, D.J.~{Schlegel}, M.A.~{Strauss}, J.~{Brinkmann},
  D.~{Finkbeiner}, M.~{Fukugita} et~al., \emph{{New York University Value-Added
  Galaxy Catalog: A Galaxy Catalog Based on New Public Surveys}},
  \href{https://doi.org/10.1086/429803}{\emph{\aj} {\bfseries 129} (2005) 2562}
  [\href{https://arxiv.org/abs/astro-ph/0410166}{{\ttfamily
  astro-ph/0410166}}].

\bibitem{2009ApJS..182..543A}
K.N.~{Abazajian}, J.K.~{Adelman-McCarthy}, M.A.~{Ag{\"u}eros}, S.S.~{Allam},
  C.~{Allende Prieto}, D.~{An} et~al., \emph{{The Seventh Data Release of the
  Sloan Digital Sky Survey}},
  \href{https://doi.org/10.1088/0067-0049/182/2/543}{\emph{\apjs} {\bfseries
  182} (2009) 543} [\href{https://arxiv.org/abs/0812.0649}{{\ttfamily
  0812.0649}}].

\bibitem{Pan2012}
D.C.~{Pan}, M.S.~{Vogeley}, F.~{Hoyle}, Y.-Y.~{Choi} and C.~{Park},
  \emph{{Cosmic voids in Sloan Digital Sky Survey Data Release 7}},
  \href{https://doi.org/10.1111/j.1365-2966.2011.20197.x}{\emph{\mnras}
  {\bfseries 421} (2012) 926}
  [\href{https://arxiv.org/abs/1103.4156}{{\ttfamily 1103.4156}}].

\bibitem{Hoyle2012}
F.~{Hoyle}, M.S.~{Vogeley} and D.~{Pan}, \emph{{Photometric properties of void
  galaxies in the Sloan Digital Sky Survey Data Release 7}},
  \href{https://doi.org/10.1111/j.1365-2966.2012.21943.x}{\emph{\mnras}
  {\bfseries 426} (2012) 3041}
  [\href{https://arxiv.org/abs/1205.1843}{{\ttfamily 1205.1843}}].

\bibitem{Park2012}
C.~{Park}, Y.-Y.~{Choi}, J.~{Kim}, I.~{Gott}, J.~Richard, S.S.~{Kim} and
  K.-S.~{Kim}, \emph{{The Challenge of the Largest Structures in the Universe
  to Cosmology}},
  \href{https://doi.org/10.1088/2041-8205/759/1/L7}{\emph{\apjl} {\bfseries
  759} (2012) L7} [\href{https://arxiv.org/abs/1209.5659}{{\ttfamily
  1209.5659}}].

\bibitem{Lee2012a}
G.-H.~{Lee}, C.~{Park}, M.G.~{Lee} and Y.-Y.~{Choi}, \emph{{Dependence of
  Barred Galaxy Fraction on Galaxy Properties and Environment}},
  \href{https://doi.org/10.1088/0004-637X/745/2/125}{\emph{\apj} {\bfseries
  745} (2012) 125} [\href{https://arxiv.org/abs/1110.1933}{{\ttfamily
  1110.1933}}].

\bibitem{Lee2012b}
G.-H.~{Lee}, J.-H.~{Woo}, M.G.~{Lee}, H.S.~{Hwang}, J.C.~{Lee}, J.~{Sohn}
  et~al., \emph{{Do Bars Trigger Activity in Galactic Nuclei?}},
  \href{https://doi.org/10.1088/0004-637X/750/2/141}{\emph{\apj} {\bfseries
  750} (2012) 141} [\href{https://arxiv.org/abs/1203.1693}{{\ttfamily
  1203.1693}}].

\bibitem{Hwang2012}
H.S.~{Hwang}, C.~{Park}, D.~{Elbaz} and Y.Y.~{Choi}, \emph{{Activity in
  galactic nuclei of cluster and field galaxies in the local universe}},
  \href{https://doi.org/10.1051/0004-6361/201117351}{\emph{\aap} {\bfseries
  538} (2012) A15} [\href{https://arxiv.org/abs/1111.1973}{{\ttfamily
  1111.1973}}].

\bibitem{Bae2014}
H.-J.~{Bae} and J.-H.~{Woo}, \emph{{A Census of Gas Outflows in Type 2 Active
  Galactic Nuclei}},
  \href{https://doi.org/10.1088/0004-637X/795/1/30}{\emph{\apj} {\bfseries 795}
  (2014) 30} [\href{https://arxiv.org/abs/1409.1580}{{\ttfamily 1409.1580}}].

\bibitem{Choi2010b}
Y.-Y.~{Choi}, C.~{Park}, J.~{Kim}, I.~{Gott}, J.~Richard, D.H.~{Weinberg},
  M.S.~{Vogeley} et~al., \emph{{Galaxy Clustering Topology in the Sloan Digital
  Sky Survey Main Galaxy Sample: A Test for Galaxy Formation Models}},
  \href{https://doi.org/10.1088/0067-0049/190/1/181}{\emph{\apjs} {\bfseries
  190} (2010) 181} [\href{https://arxiv.org/abs/1005.0256}{{\ttfamily
  1005.0256}}].

\bibitem{dunkley2009}
J.~{Dunkley}, E.~{Komatsu}, M.R.~{Nolta}, D.N.~{Spergel}, D.~{Larson},
  G.~{Hinshaw} et~al., \emph{{Five-Year Wilkinson Microwave Anisotropy Probe
  Observations: Likelihoods and Parameters from the WMAP Data}},
  \href{https://doi.org/10.1088/0067-0049/180/2/306}{\emph{\apjs} {\bfseries
  180} (2009) 306} [\href{https://arxiv.org/abs/0803.0586}{{\ttfamily
  0803.0586}}].

\bibitem{jiang2008}
C.Y.~{Jiang}, Y.P.~{Jing}, A.~{Faltenbacher}, W.P.~{Lin} and C.~{Li}, \emph{{A
  Fitting Formula for the Merger Timescale of Galaxies in Hierarchical
  Clustering}}, \href{https://doi.org/10.1086/526412}{\emph{\apj} {\bfseries
  675} (2008) 1095} [\href{https://arxiv.org/abs/0707.2628}{{\ttfamily
  0707.2628}}].

\bibitem{Zehavi2011}
I.~{Zehavi}, Z.~{Zheng}, D.H.~{Weinberg}, M.R.~{Blanton}, N.A.~{Bahcall},
  A.A.~{Berlind} et~al., \emph{{Galaxy Clustering in the Completed SDSS
  Redshift Survey: The Dependence on Color and Luminosity}},
  \href{https://doi.org/10.1088/0004-637X/736/1/59}{\emph{\apj} {\bfseries 736}
  (2011) 59} [\href{https://arxiv.org/abs/1005.2413}{{\ttfamily 1005.2413}}].

\bibitem{Zehavi2002}
I.~{Zehavi}, M.R.~{Blanton}, J.A.~{Frieman}, D.H.~{Weinberg}, H.J.~{Mo},
  M.A.~{Strauss} et~al., \emph{{Galaxy Clustering in Early Sloan Digital Sky
  Survey Redshift Data}}, \href{https://doi.org/10.1086/339893}{\emph{\apj}
  {\bfseries 571} (2002) 172}
  [\href{https://arxiv.org/abs/astro-ph/0106476}{{\ttfamily
  astro-ph/0106476}}].

\bibitem{Guo2012}
H.~{Guo}, I.~{Zehavi} and Z.~{Zheng}, \emph{{A New Method to Correct for Fiber
  Collisions in Galaxy Two-point Statistics}},
  \href{https://doi.org/10.1088/0004-637X/756/2/127}{\emph{\apj} {\bfseries
  756} (2012) 127} [\href{https://arxiv.org/abs/1111.6598}{{\ttfamily
  1111.6598}}].

\bibitem{Reid2014}
B.A.~{Reid}, H.-J.~{Seo}, A.~{Leauthaud}, J.L.~{Tinker} and M.~{White},
  \emph{{A 2.5 per cent measurement of the growth rate from small-scale
  redshift space clustering of SDSS-III CMASS galaxies}},
  \href{https://doi.org/10.1093/mnras/stu1391}{\emph{\mnras} {\bfseries 444}
  (2014) 476} [\href{https://arxiv.org/abs/1404.3742}{{\ttfamily 1404.3742}}].

\bibitem{morey1984}
L.C.~Morey and A.~Agresti, \emph{The measurement of classification agreement:
  An adjustment to the rand statistic for chance agreement}, {\emph{Educational
  and Psychological Measurement} {\bfseries 44} (1984) 33}.

\bibitem{hubert1985}
L.~Hubert and P.~Arabie, \emph{Comparing partitions journal of classification 2
  193--218}, {\emph{Google Scholar} (1985) 193}.

\bibitem{rand1971}
W.M.~Rand, \emph{Objective criteria for the evaluation of clustering methods},
  {\emph{Journal of the American Statistical association} {\bfseries 66} (1971)
  846}.

\bibitem{Dall2002}
J.~{Dall} and M.~{Christensen}, \emph{{Random geometric graphs}},
  \href{https://doi.org/10.1103/PhysRevE.66.016121}{\emph{\pre} {\bfseries 66}
  (2002) 016121} [\href{https://arxiv.org/abs/cond-mat/0203026}{{\ttfamily
  cond-mat/0203026}}].

\bibitem{peebles1993}
P.J.E.~{Peebles}, \emph{{Principles of Physical Cosmology}}, Princeton
  university press (1993),
  \href{https://doi.org/10.1515/9780691206721}{10.1515/9780691206721}.

\bibitem{peacock1999}
J.A.~{Peacock}, \emph{{Cosmological Physics}}, Cambridge university press
  (1999).

\bibitem{Robotham2011}
A.S.G.~Robotham, P.~Norberg, S.P.~Driver, I.K.~Baldry, S.P.~Bamford,
  A.M.~Hopkins et~al., \emph{{Galaxy and Mass Assembly (GAMA): the GAMA galaxy
  group catalogue (G3Cv1)}},
  \href{https://doi.org/10.1111/j.1365-2966.2011.19217.x}{\emph{Monthly Notices
  of the Royal Astronomical Society} {\bfseries 416} (2011) 2640}
  [\href{https://arxiv.org/abs/https://academic.oup.com/mnras/article-pdf/416/4/2640/2973612/mnras0416-2640.pdf}{{\ttfamily
  https://academic.oup.com/mnras/article-pdf/416/4/2640/2973612/mnras0416-2640.pdf}}].

\bibitem{Yang2021}
X.~Yang, H.~Xu, M.~He, Y.~Gu, A.~Katsianis, J.~Meng et~al., \emph{An extended
  halo-based group/cluster finder: Application to the desi legacy imaging
  surveys dr8}, \href{https://doi.org/10.3847/1538-4357/abddb2}{\emph{The
  Astrophysical Journal} {\bfseries 909} (2021) 143}.

\end{thebibliography}\endgroup
%===============================================================================

\end{document}